\documentclass[a4paper,12pt,titlepage,twoside]{article}
\usepackage{graphicx}
\usepackage{graphics}
\usepackage{color}
\usepackage{a4}
\usepackage{setspace}

\title{Multilayer Adsorption of Polyatomic Species on Homogeneous and Heterogeneous Surfaces}

\author{F. O. S\'anchez-Varretti$^1$, G. D. Garc\'{\i}a$^1$, A. J. Ramirez-Pastor$^2$ and F. Rom\'a$^{2,3}$
\\\\ $^1$Universidad Tecnol\'ogica Nacional, Regional San Rafael, \\
 Gral. Urquiza 314, 5600, San Rafael, Mendoza, Argentina \\
$^2$Departamento de F\'{\i}sica, Universidad Nacional de San Luis,
\\ Chacabuco 917, San Luis D5700BWS, Argentina\\
$^3$ Centro At{\'{o}}mico Bariloche, San Carlos de Bariloche, \\
R\'{\i}o Negro R8402AGP , Argentina}

\begin{document}

\maketitle

\begin{abstract}
In this work we study the multilayer adsorption of polyatomic
species on homogeneous and heterogeneous bivariate surfaces. A new
approximate analytic isotherm is obtained and validated by
comparing with Monte Carlo simulation. Then, we use the well-known
Brunauer-Emmet-Teller's (BET) approach to analyze these isotherms
and to estimate the monolayer volume, $v_\mathrm{m}$.  The results
show that the value of the $v_\mathrm{m}$ obtained in this way
depends strongly on adsorbate size and surface topography. In all
cases, we find that the use of the BET equation leads to an
underestimate of the true monolayer capacity.
\end{abstract}

\section{Introduction}

The theoretical description of multilayer adsorption is a
long-standing important problem in surface science that does not
have a general solution. \cite{Gregg1991,Rudzinski1992} Mainly
this is due to the fact that the structure of the different layers
differs from that in contact with the solid surface (first layer).
At high coverage (multilayer region), it is expected that the
adsorption process is well described by the slab theory of
Frenkel, Halsey and Hill. \cite{Frenkel1946,Halsey1948,Hill1952}
In this approach it is assumed that the higher layers retain the
structure of the bulk liquid, and only its free energy changes
gradually as one goes away from the solid surface.  On the other
hand, at low coverage, it is more appropriate to use the
Brunauer-Emmet-Teller's (BET) isotherm, \cite{Brunauer1938} where
the crystal-like structure of the surface is considered.  In the
BET theory it is assumed that the molecules are localized in sites
and that the adsorption in the first layer is different from the
remaining ones.

The BET equation is one of the most widely used isotherms. The
approach discards such things as the polyatomic character of the
adsorbate, the interaction between the admolecules and the surface
heterogeneities.  Then, with the purpose of including a more
complex situation, numerous generalizations of the BET theory have
been proposed. \cite{Rudzinski1992,Hill1962}  Nevertheless, the
simplicity of the BET isotherm has made it very popular for
practical purposes.  In fact, by fitting an experimental isotherm
with the BET equation, in many cases it is possible to predict the
monolayer volume (or monolayer capacity) of the solid surface with
an error not bigger than the $20 \%$. \cite{Gregg1991}  This
surprising result is attributed to compensation effects arisen as
consequence of having carried out many approximations.
\cite{Gregg1991}

By means of numerical experiments, Walker and Zettlemoyer
\cite{Walker1948} have shown that the conventional BET theory
predicts a monolayer volume smaller than the real value, when
heterogeneous adsorption isotherms are analyzed.  A similar result
was obtained by Cort\'es and Araya, \cite{Cortes1987} when
considering a Gaussian distribution of adsorption energies.  The
authors showed that the estimated values of the monolayer volume
from the BET equation diminishes with increasing degree of surface
heterogeneity (the width of the distribution of adsorption
energies).

More recently, Nikitas \cite{Nikitas1996} has arrived to similar
conclusions by considering both, surface heterogeneity and
polyatomic character of the adsorbate.  In ref~\cite{Nikitas1996},
by using an extension of the Flory-Huggins polymer solution
theory, \cite{Hill1962} the multilayer adsorption of polyatomic
species was studied  over a random heterogeneous surface. By
particular cases analysis, the author concludes that one can
obtain an underestimation of the true monolayer capacity of the
order of $25\%$, when the adsorbate occupies more than one lattice
site. This underestimation is bigger, if an heterogeneous surface
is considered.

In this work, we study how the monolayer volume predicted by BET
equation differs from its real value when considering both the
adsorbate size and the surface topography, i. e. the space
distribution of the adsorption energies over the solid surface. In
particular, we consider the multilayer adsorption of polyatomic
species on one-dimensional (1D) and two-dimensional (2D)
homogeneous and heterogeneous bivariate surfaces.  In each case,
approximate analytic isotherms are built and validated by
comparing with Monte Carlo simulation. Then, we estimate the
monolayer volume, by analyzing these isotherms with the
conventional BET theory.

The paper is structured as follows.  In section 2, we present the
lattice-gas model. Next, in sections 3 and 4, the multilayer
adsorption isotherms for homogeneous and heterogeneous surfaces
are obtained.  The dependence of the monolayer volume on the
surface topography and the adsorbate size is presented in section
5. Finally, conclusions are drawn in section 6.

\section{The Lattice-Gas Model}

A simple lattice-gas model to describe the multilayer adsorption
of polyatomic molecules on homogeneous surface has been recently
proposed. \cite{Riccardo2002,Roma2005} The surface is modeled by a
regular lattice of $M$ sites, all with the same adsorption energy
$\varepsilon$, and the adsorbate is represented by $k$-mers
(linear particles that have $k$ identical units). A $k$-mer
adsorbed on the surface occupies $k$ sites of the lattice with an
energy $k \varepsilon$ and can arrange in many configurations.
This property is called adsorption with multisite occupancy. On
the other hand, for higher layers, the adsorption of a $k$-mer is
exactly onto an already absorbed one, with an adsorption energy of
$k U$. Thus, the monolayer structure reproduces in the remaining
layers. This phenomenon is called {\it pseudomorphism} and, for
example, is observed experimentally in the adsorption of straight
chain saturated hydrocarbon molecules. \cite{Somorjai1979}
Finally, following the spirit of the BET theory, no lateral
interactions are considered and only interactions among the layers
are introduced. Figure 1 in ref~\cite{Riccardo2002} shows a
snapshot representing this lattice-gas model.

We modify the model to consider the adsorption on a heterogeneous
substrate: now the adsorption energy $\varepsilon_i$ depends on
each site $i$ of the surface. Then, the Hamiltonian of the system
is
\begin{equation}
H=\left( N-N_{\mathrm{m}} \right)k U + \sum_{i=1}^{M} \sigma_i
\varepsilon_i, \label{ham1}
\end{equation}
where $N$ is the total number of $k$-mers, $\sigma_i$ the
occupation variable which can take the values 0 if the
corresponding site is empty or 1 if the site is occupied, and
$N_{\mathrm{m}}$ is the number of $k$-mers on the surface
(monolayer),
\begin{equation}
N_{\mathrm{m}}= \frac{1}{k} \sum_{i=1}^{M} \sigma_i .
\end{equation}
The Hamiltonian in eq~(\ref{ham1}) can be rewritten as
\begin{equation}
H=N k U + \sum_{i=1}^{M} \sigma_i \left( \varepsilon_i - U \right)
. \label{ham2}
\end{equation}
In the following sections, we will look for analytic solutions of
the model in 1D and 2D, for homogeneous and heterogeneous
surfaces.

\section{Multilayer Adsorption on Homogeneous
Surfaces}

For the previous model, a simple analytic expression of the
multilayer adsorption isotherm can be obtained in few particular
cases. If the surface is homogeneous ($\varepsilon_i=\varepsilon$
for all $i$) and $k=1$, it is easy to demonstrate
\cite{Gregg1991,Rudzinski1992,Riccardo2002} that
\begin{equation}
\theta = \frac{1}{\left( 1- p/p_0 \right) } \frac{c p/p_0}{\left[
1+ \left(c-1 \right) p/p_0 \right]}. \label{BET}
\end{equation}
Here, $\theta = k N/M$ is the total coverage, $p$ is the pressure,
 $p_0$ is the saturation pressure of the bulk liquid and $c$ is a constant defined as
\begin{equation}
c=\exp \left[ - \beta k \left( \varepsilon - U \right) \right],
\label{c}
\end{equation}
where $\beta= 1/ k_B T$ is the inverse temperature (being $k_B$
the Boltzmann constant and $T$ the absolute temperature).
eq~(\ref{BET}) is the well-known BET isotherm \cite{Brunauer1938}
and can be applied to systems in any dimension.

In the case of $k=2$, it is only possible to obtain an exact
solution in 1D \cite{Riccardo2002}
\begin{equation}
\theta = \frac{1}{\left( 1- p/p_0 \right)}  \left\{ 1 - \left[
\frac{1-p/p_0}{1+ \left( 4c-1 \right)p/p_0} \right]^{1/2}
\right\}. \label{dimer}
\end{equation}
Equation~(\ref{dimer}) is the exact dimer isotherm for the
multilayer adsorption on 1D homogeneous surface.  As it has been
previously shown, \cite{Riccardo2002} the values of the monolayer
volume predicted by eqs~(\ref{BET}) and (\ref{dimer}) are
different: if a dimer isotherm is analyzed, the value of the
monolayer capacity arising from using BET characterization is
smaller than the real one.

In general, the multilayer isotherm corresponding to the model of
eq~(\ref{ham2}) cannot be expressed by one equation only. To
describe the isotherm, it is necessary to give two functions.
\cite{Nikitas1996,Roma2005} Let us assume that an analytical
expression of the monolayer adsorption isotherm is known, $\exp
\left ( \mu \beta \right) = \lambda$, being $\mu$ the chemical
potential and $\lambda$ (the fugacity) a function of the monolayer
coverage $\theta_{\mathrm{m}}= k N_{\mathrm{m}}/M$. Then, the
following equations can be deduced \cite{Roma2005}:
\begin{equation}
\frac{p}{p_0} = \frac{\lambda \left ( \theta_{\mathrm{m}}
\right)}{c + \lambda \left ( \theta_{\mathrm{m}} \right)}
\label{pre}
\end{equation}
and
\begin{equation}
\theta = \frac{\theta_{\mathrm{m}}}{\left( 1+p/p_0 \right)}.
\label{cub}
\end{equation}
Equations~(\ref{pre}) and (\ref{cub}) constitute the multilayer
adsorption isotherm.  By using the surface coverage ($0 \leq
\theta_{\mathrm{m}} \leq 1$) as a parameter, we can calculate the
relative pressure from eq~(\ref{pre}). Then, the values of
$\theta_{\mathrm{m}}$ and $p/p_0$ are introduced in eq~(\ref{cub})
and the total coverage is obtained.

\begin{figure}[t]
\centering
\includegraphics[width=6.5cm,clip=true]{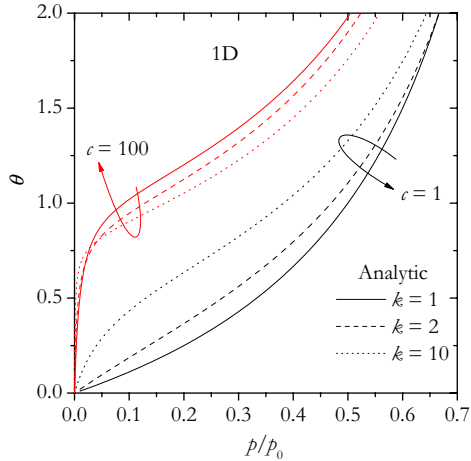}
\caption{\label{figure1} Exact 1D isotherms for $k=1$, 2 and 10,
and two values of $c$ as indicated.}
\end{figure}

Following the previous scheme, it is possible to obtain the exact
multilayer isotherm for $k$-mers in 1D homogeneous surfaces.
\cite{Roma2005} We start from the exact monolayer isotherm
\cite{Ramirez-Pastor1999}
\begin{equation}
\lambda = \frac{\theta_{\mathrm{m}} \left[ 1- \frac{\left( k-1
\right)}{k} \theta_{\mathrm{m}} \right]^{k-1}}{k \left(
1-\theta_{\mathrm{m}}\right)^k}.\label{mon1D}
\end{equation}
Then, eq~(\ref{pre}) can be written as
\begin{equation}
\frac{p}{p_0} = \frac{\theta_{\mathrm{m}} \left[ 1- \frac{\left(
k-1 \right)}{k} \theta_{\mathrm{m}} \right]^{k-1}}{ k c \left(
1-\theta_{\mathrm{m}}\right)^k + \theta_{\mathrm{m}} \left[ 1-
\frac{\left( k-1 \right)}{k} \theta_{\mathrm{m}} \right]^{k-1}}.
\label{pre1D}
\end{equation}
Equations~(\ref{pre1D}) and (\ref{cub}) represent the exact
solution of the 1D model.   In particular, for $k=1$ and $k=2$, it
is possible to solve these equations to obtain single expressions
of the multilayer isotherms, eqs~(\ref{BET}) and (\ref{dimer}),
respectively. Figure~\ref{figure1} shows the exact 1D isotherms
for $k=1$, 2 and 10, and two values of $c$.

Also, the previous scheme can be used to obtain an accurate
approximation for multilayer adsorption on 2D substrates
accounting multisite occupancy. In this case, we use the
semi-empirical (SE) monolayer adsorption isotherm
\cite{Roma2006,Riccardo2006}
\begin{equation}
\lambda = \frac{2 \theta_{\mathrm{m}} \left[ 1- \frac{\left( k-1
\right)}{k} \theta_{\mathrm{m}} \right]^{\left(k-1 \right)
\theta_\mathrm{m}} \left[ 1- \frac{2 \left( k-1 \right)}{\gamma k}
\theta_{\mathrm{m}} \right]^{\left(k-1 \right) \left( 1-
\theta_\mathrm{m}\right)}} {\gamma k \left(
1-\theta_{\mathrm{m}}\right)^k}, \label{SE}
\end{equation}
where $\gamma$ is the connectivity of the lattice. It has been
shown that eq~(\ref{SE}) is a very good approximation for
representing multisite-occupancy adsorption in the monolayer
regime. \cite{Roma2006,Riccardo2006} Then, by using
eqs~(\ref{pre}) and (\ref{SE}) we obtain
\begin{equation}
\frac{p}{p_0} = \frac{2 \theta_{\mathrm{m}} \left[ 1- \frac{\left(
k-1 \right)}{k} \theta_{\mathrm{m}} \right]^{\left(k-1 \right)
\theta_\mathrm{m}} \left[ 1- \frac{2 \left( k-1 \right)}{\gamma k}
\theta_{\mathrm{m}} \right]^{\left(k-1 \right) \left( 1-
\theta_\mathrm{m}\right)}}{\gamma k c \left(
1-\theta_{\mathrm{m}}\right)^k + 2 \theta_{\mathrm{m}} \left[ 1-
\frac{\left( k-1 \right)}{k} \theta_{\mathrm{m}}
\right]^{\left(k-1 \right) \theta_\mathrm{m}} \left[ 1- \frac{2
\left( k-1 \right)}{\gamma k} \theta_{\mathrm{m}}
\right]^{\left(k-1 \right) \left( 1- \theta_\mathrm{m}\right)}}.
\label{pre2D}
\end{equation}
Note that, for $\gamma = 2$, the SE isotherm is identical to the
eq~(\ref{mon1D}).  Therefore, eqs~(\ref{cub}) and (\ref{pre2D})
represent the general solution of the problem of multilayer
adsorption in homogeneous surfaces with multisite occupancy: for
$\gamma = 2$ (1D) this isotherm is exact, but is approximate for
$\gamma > 2$. In addition, in any dimension, the exact isotherm
for $k=1$ (BET equation) can also be obtained from
eqs~(\ref{cub}) and (\ref{pre2D}), but always with $\gamma = 2$
(these equations with $k=1$ and $\gamma > 2$ are erroneous,
because do not provide the BET isotherm).

\begin{figure}[t]
\centering
\includegraphics[width=6.5cm,clip=true]{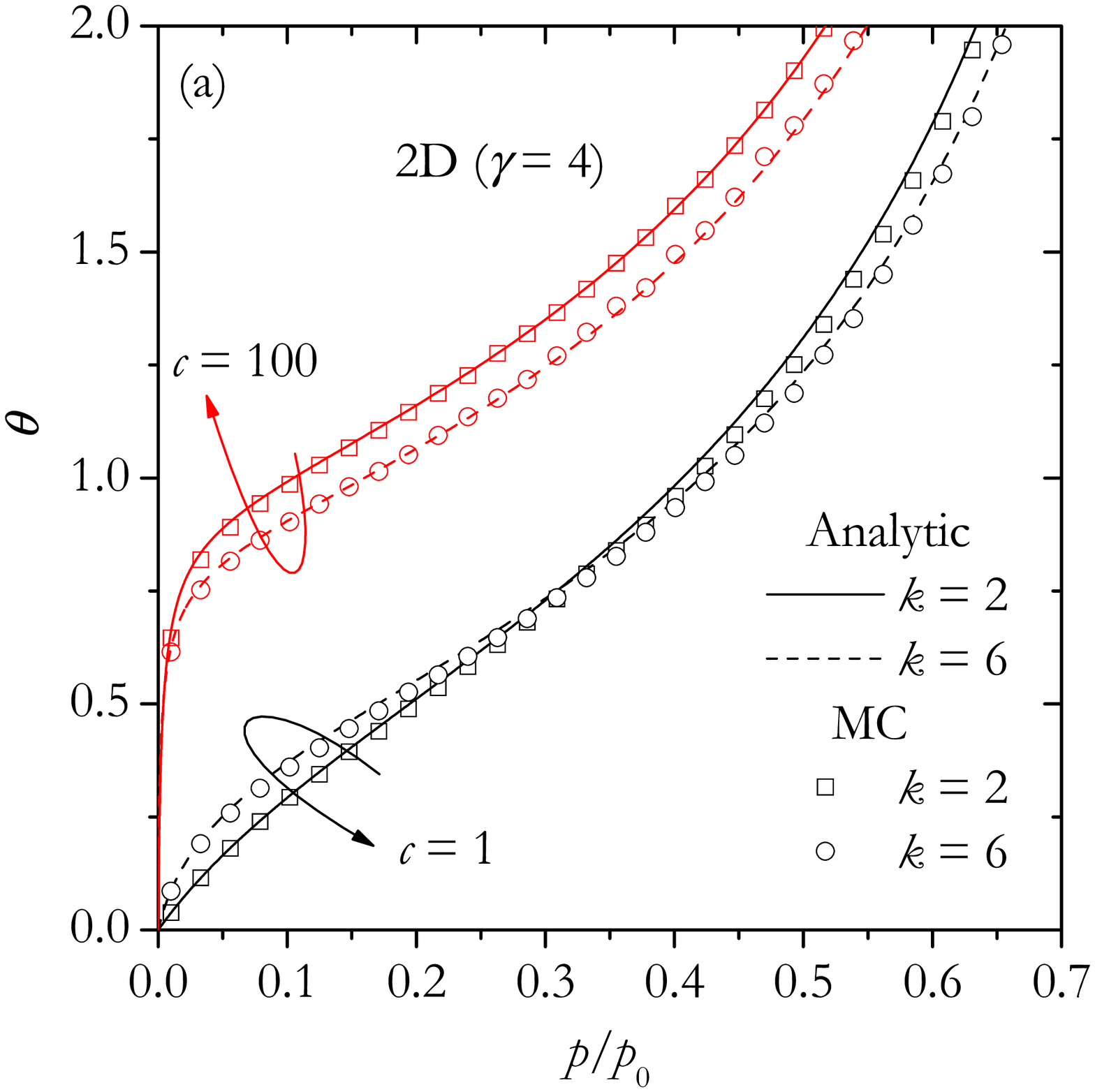}
\includegraphics[width=6.5cm,clip=true]{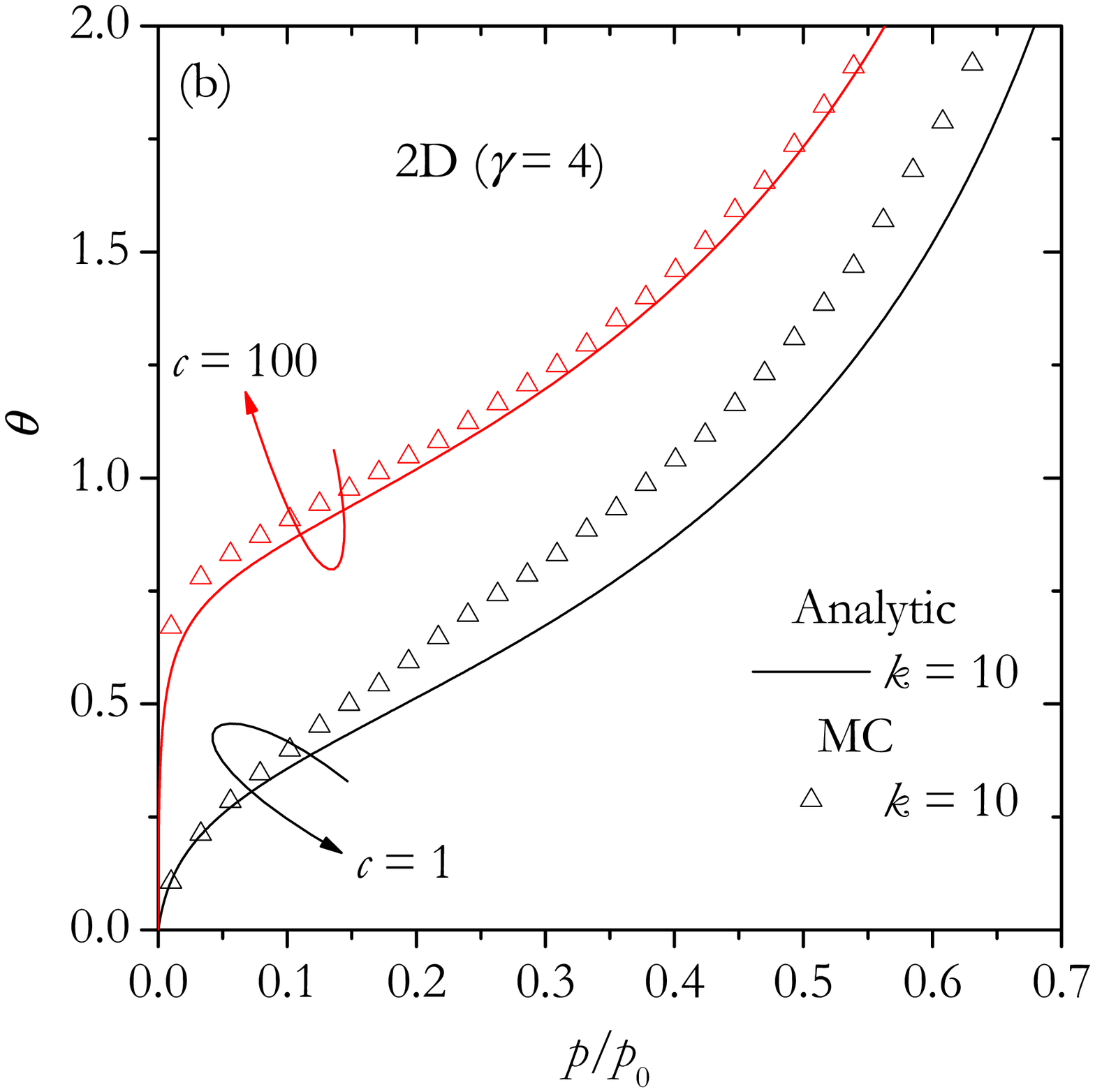}
\caption{\label{figure2} Comparison between analytic and simulated
adsorption isotherms for 2D lattices and two different values of
$c$ as indicated. (a) $k=2$ and 6, (b) $k=10$.  In all cases, we
have used $t=10^5$ MCSs.}
\end{figure}

In order to test the 2D approximation, we have compared the
analytic multilayer isotherm with results of Monte Carlo (MC)
simulation.  The algorithm used is described in
ref~\cite{Roma2005}. Here, the equilibrium state is reproduced
after discarding a number $t$ of MC steps (MCSs). Then, the mean
value of the total coverage is obtained as,
\begin{equation}
\theta=\frac{k \langle N \rangle}{M},
\end{equation}
where the average $\langle N \rangle$ is calculated over another
$t$ successive MCSs (the total number of MCSs is $2t$).  The
computational simulations were developed for a square lattice
($\gamma = 4$) of linear size $L$ ($M=L\times L$).  For each value
of $k$, we choose $L=20k$.  For these lattice sizes (proportional
to $k$), we have verified that finite-size effects are negligible.

Figure~\ref{figure2}a shows a comparison between the analytic
isotherm [given by eqs~(\ref{cub}) and (\ref{pre2D})] and the MC
results, for $k=2$ and 6 and two values of $c$.  As we can see,
 the agreement is very good for the parameters used in the figure.
 On the other hand, the accuracy of the analytic isotherm diminishes as
$k$ increases.  Figure~\ref{figure2}b shows this effect for
$k=10$. Also, in this figure, we can appreciate that the
difference between the analytic and the numerical isotherm
diminishes as $c$ is increased.

\section{Multilayer Adsorption on Heterogeneous
Surfaces}

In the previous section, we have obtained the multilayer isotherm
from the monolayer isotherm.  It is possible demonstrate that in
general, the formalism allows to establish this connection only if
$(1)$ pseudomorphism is present and $(2)$ no lateral interactions
between the molecules in the multilayer regime are considered. In
fact, eqs~(\ref{pre}) and (\ref{cub}) still hold if the particles
in the monolayer interact among them and with the solid surface.
Although we could use this formalism to determine the multilayer
adsorption isotherm for a given surface heterogeneity (for which
it would be necessary to have an appropriate monolayer isotherm),
we have chosen to use a different strategy.

We start here from the integral representation of the adsorption
multilayer isotherm \cite{Rudzinski1992}
\begin{equation}
\theta = \int \chi(\varepsilon) \theta_\mathrm{loc}(\varepsilon) d
\varepsilon, \label{integral}
\end{equation}
where $\theta_\mathrm{loc}(\varepsilon)$ represents the local
adsorption multilayer isotherm corresponding to an adsorptive site
of energy $\varepsilon$ and $\chi(\varepsilon)$ is the adsorptive
energy distribution which characterizes the surface heterogeneity
(as before, the total and the local coverage depend on $p$ and
$T$).

It should be noticed that eq~(\ref{integral}) is strictly and
generally valid only for noninteracting monomers ($k=1$), which is
a quite unrealistic case. If adsorbed particles occupy more than
one site (multisite occupancy) or interact with each other, then
the local coverage at a point with a given adsorptive energy
depends on the local coverage on neighbor points with different
adsorptive energies and, in general, eq~(\ref{integral}) should be
replaced by a much more complex expression.
\cite{Riccardo1992,Bulnes2001}

Nevertheless, in some situations eq~(\ref{integral}) represents a
good approximation of the adsorption isotherm (see below). For a
lattice-gas model of $k$-mers, we can generalize this equation as
\begin{equation}
\theta = \sum_{s} \theta_\mathrm{loc}(E_s). \label{isoh}
\end{equation}
In the last equation, the sum extends over all possible
configurations of a single $k$-mer in an empty lattice, and $E_s$
is the adsorption energy of each one of them. Note that the values
of $E_s$ depend, among other things, on the energy distribution
$\chi(\varepsilon)$, the surface topography and the number $k$.

In following sections, we will study the multilayer adsorption on
1D and 2D heterogeneous surfaces. As local isotherm, we will use
eqs~(\ref{cub}) and (\ref{pre2D}). Then, we will compare the
multilayer adsorption isotherm obtained by using eq~(\ref{isoh})
and the calculated with MC simulation.

\subsection{Adsorption on 1D Heterogeneous
Surfaces}

As we said in the introduction, the heterogeneity is modeled by
two kinds of sites (bivariate surface): {\it strong} sites with
adsorption energy $\varepsilon_1$ and {\it weak} sites with
adsorption energy $\varepsilon_2$ ($\varepsilon_1 <
\varepsilon_2$).  In 1D, where the surface is represented by a
chain of sites with periodic boundary conditions, these sites form
patches of size $l$ $(l = 1, 2, 3,\cdots)$, which are spatially
distributed (topography) in a deterministic alternate way.

The number of possible configurations of a single $k$-mer in an
empty lattice is $M$.  However, due to periodicity,
eq~(\ref{isoh}) has only $2l$ terms (with many of them having the
same adsorption energy). Then, the multilayer isotherm is
approximated as
\begin{equation}
\theta = \frac{1}{2l} \sum_{i=1}^{2l} \theta_\mathrm{loc} \left(
c_i \right). \label{isoh1D}
\end{equation}
Each term corresponds to an effective value of $c$ given by
\begin{equation}
c_i=\exp \left[ - \beta \left( E_i - k U \right) \right],
\end{equation}
where $E_i$ is the adsorption energy.  This value of $c_i$ can
also be expressed as function of $c_1$ and $c_2$, the values of
$c$ for homogeneous surfaces given by eq~(\ref{c}) and whose
adsorption energies are $\varepsilon_1$ and $\varepsilon_2$,
respectively.  If the $i$-th term in eq~(\ref{isoh1D}) corresponds
to a $k$-mer with $k_1$ units located over strong sites and $k_2$
units located over weak sites, then the adsorption energy is
$E_i=k_1 \varepsilon_1+ k_2 \varepsilon_2$, and
\begin{equation}
c_i=\left(c_1^{k_1} c_2^{k_2} \right)^{1/k}.
\end{equation}

As mentioned previously, we use eqs~(\ref{cub}) and (\ref{pre2D})
with $\gamma = 2$ as local isotherm. Note that eq~(\ref{isoh1D})
is a sum of local isotherms with different values of $c$, but {\it
at the same relative pressure}. Then, in most of the cases it is
necessary to be careful: although, for each local isotherm the
surface coverage should be used as a parameter, it is not possible
to use this as common parameter.  In fact, eq~(\ref{pre}) shows
that for a fixed value of $p/p_0$, the surface coverage depends on
$c$, $\theta_{\mathrm{m}} (c_i)$.

Now, we analyze two simple cases.  On one hand, eq~(\ref{isoh1D})
is exact for $k=1$ and can be obtained as the semisum of two BET
isotherms,
\begin{equation}
\theta = \frac{p/p_0}{2 \left( 1- p/p_0 \right)} \left\{ \frac{c_1
}{\left[ 1+ \left(c_1-1 \right) p/p_0 \right]} + \frac{c_2
}{\left[ 1+ \left(c_2-1 \right) p/p_0 \right]}\right\}.
\label{isoh1Dk1}
\end{equation}
In this case, eq~(\ref{isoh1Dk1}) does not depend on $l$ and,
consequently, the multilayer adsorption isotherm is the same for
all topography.

On the other hand, eq~(\ref{isoh1D}) has three different terms for
$k=2$, being each one of them a dimer isotherm eq~(\ref{dimer})
with a particular value of $c$. Thus, for $k=2$, the multilayer
adsorption isotherm is
\begin{eqnarray}
\theta &=& \left( \frac{l-1}{2l} \right) \frac{1}{\left( 1- p/p_0
\right)} \left\{ 1 - \left[ \frac{1-p/p_0}{1+ \left( 4c_1-1
\right)p/p_0} \right]^{1/2}
\right\} + \nonumber\\
&+& \left( \frac{1}{l} \right) \frac{1}{\left( 1- p/p_0 \right)}
\left\{ 1 - \left[ \frac{1-p/p_0}{1+ \left( 4 \sqrt{c_1 c_2} -1
\right)p/p_0}
\right]^{1/2} \right\} + \nonumber\\
&+& \left( \frac{l-1}{2l} \right) \frac{1}{\left( 1- p/p_0
\right)} \left\{ 1 - \left[ \frac{1-p/p_0}{1+ \left( 4c_2-1
\right)p/p_0} \right]^{1/2} \right\}. \label{isoh1Dk2}
\end{eqnarray}
The first [third] term in the RHS of eq~(\ref{isoh1Dk2})
represents the adsorption within a strong [weak] patch, on a pair
of sites (1,1) [(2,2)], with $c_1$ [$c_2$]. There are $(l-1)$
configurations of this for each patch. The remaining term of
eq~(\ref{isoh1Dk2}) corresponds to a dimer isotherm with
adsorption energy $E= \varepsilon_1+ \varepsilon_2$ ($c=\sqrt{c_1
c_2}$). There are only two configurations with this energy for
pairs of sites (1,2) or (2,1). Contrary to eq~(\ref{isoh1Dk1}),
eq~(\ref{isoh1Dk2}) depends on $l$ and the dimer isotherm {\it
sees} the topography.

\begin{figure}[t]
\centerline{
\includegraphics[width=6.5cm,clip=true]{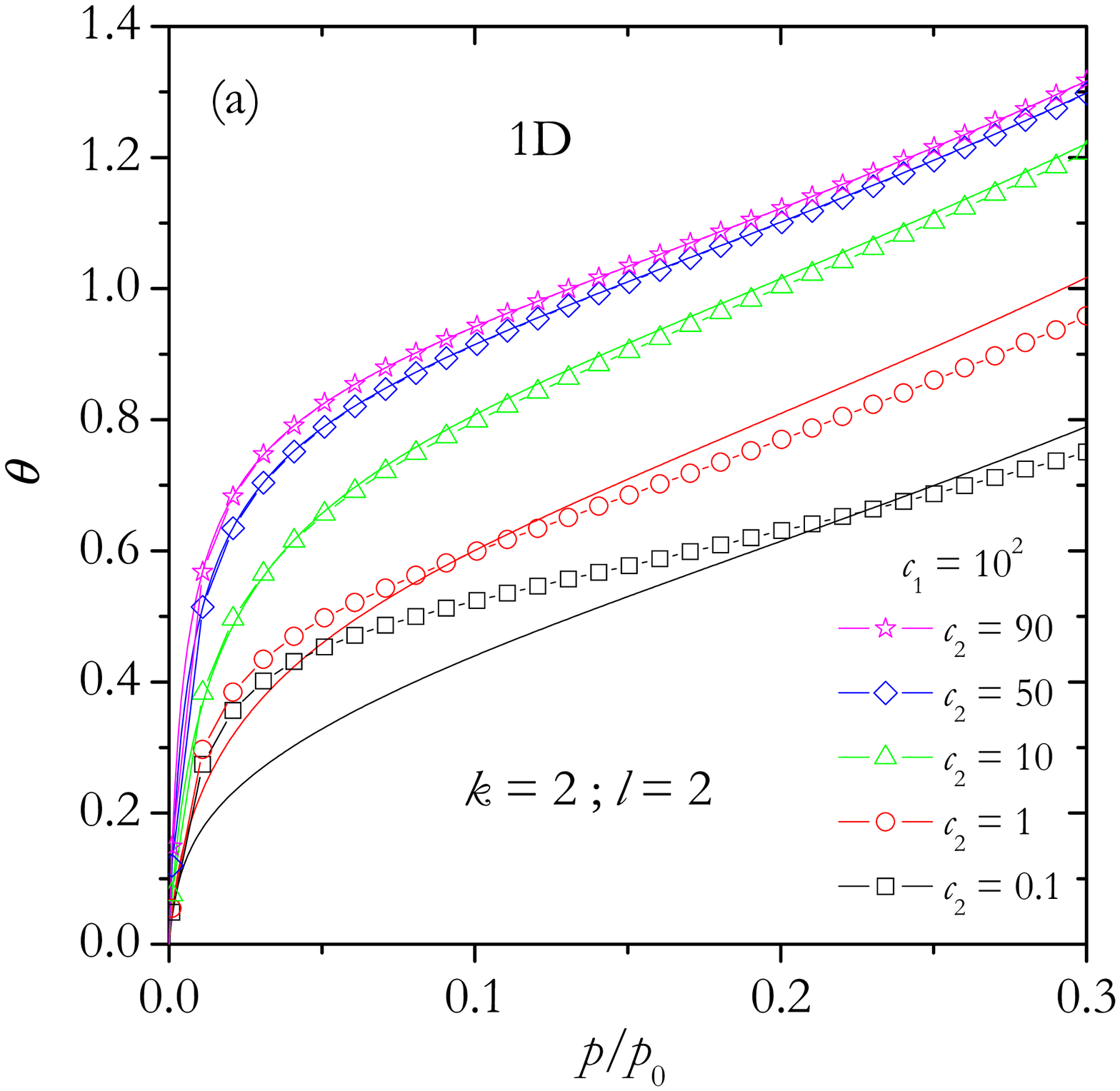}
\includegraphics[width=6.5cm,clip=true]{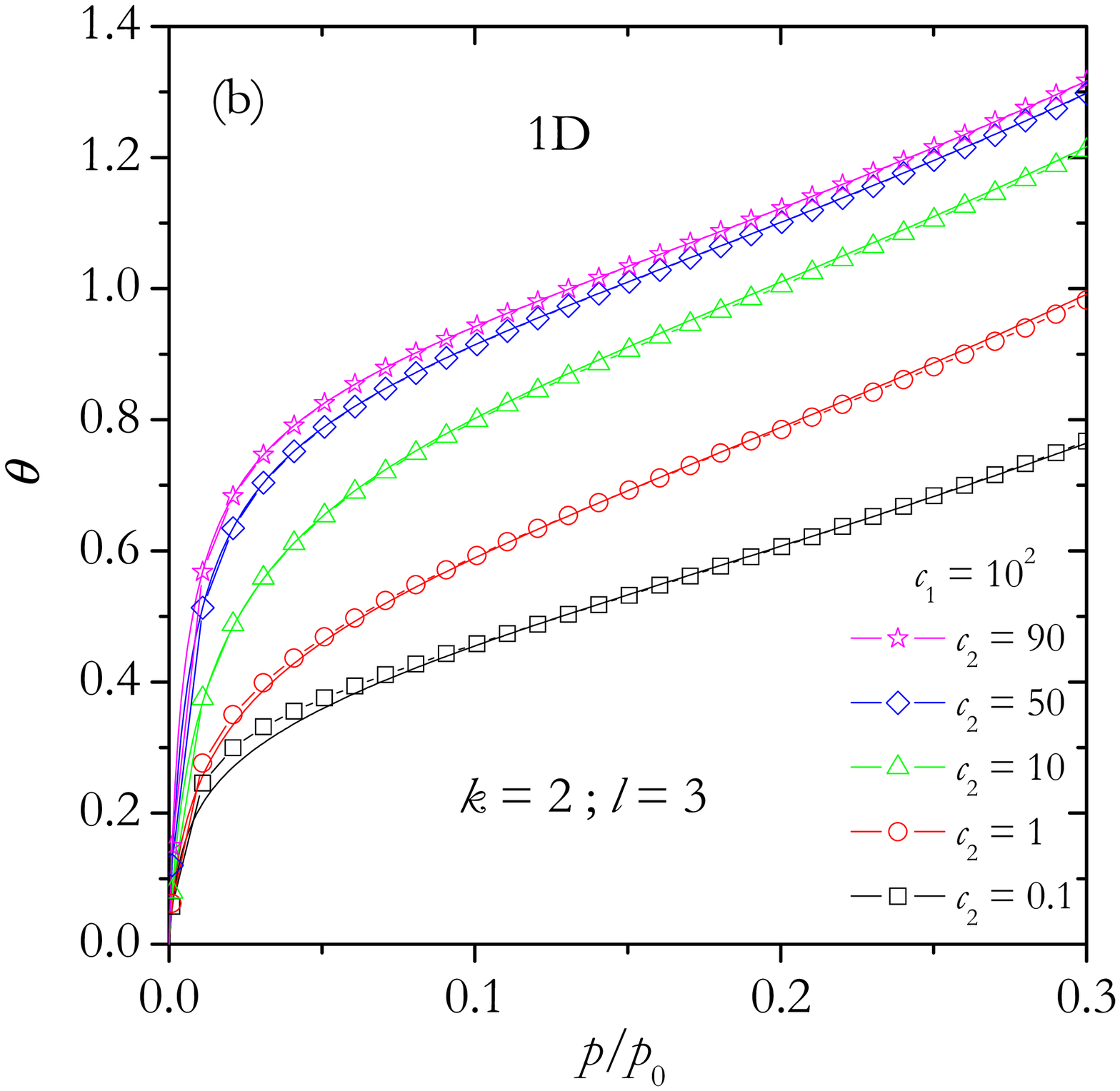}}
\centerline{
\includegraphics[width=6.5cm,clip=true]{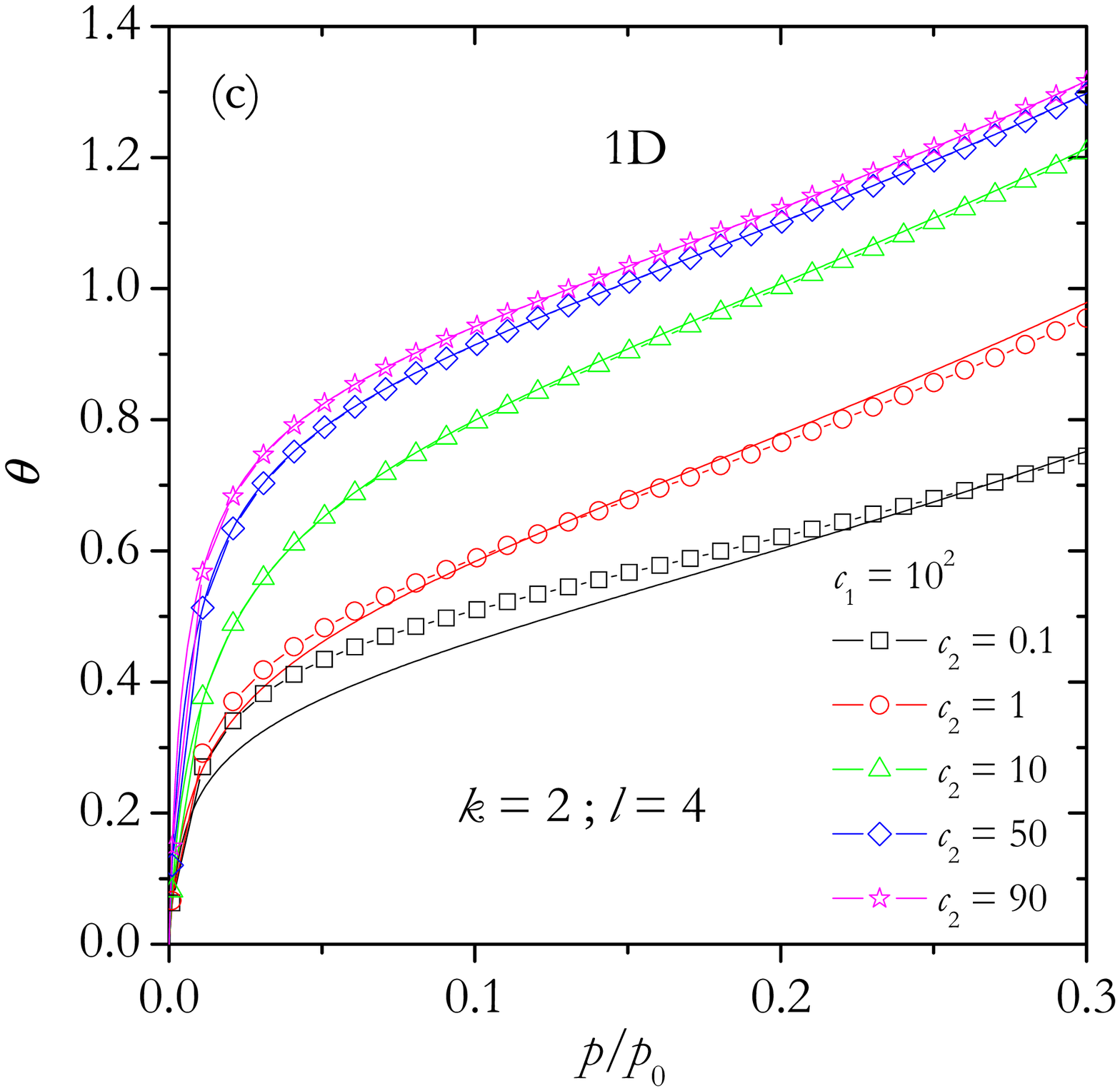}
\includegraphics[width=6.5cm,clip=true]{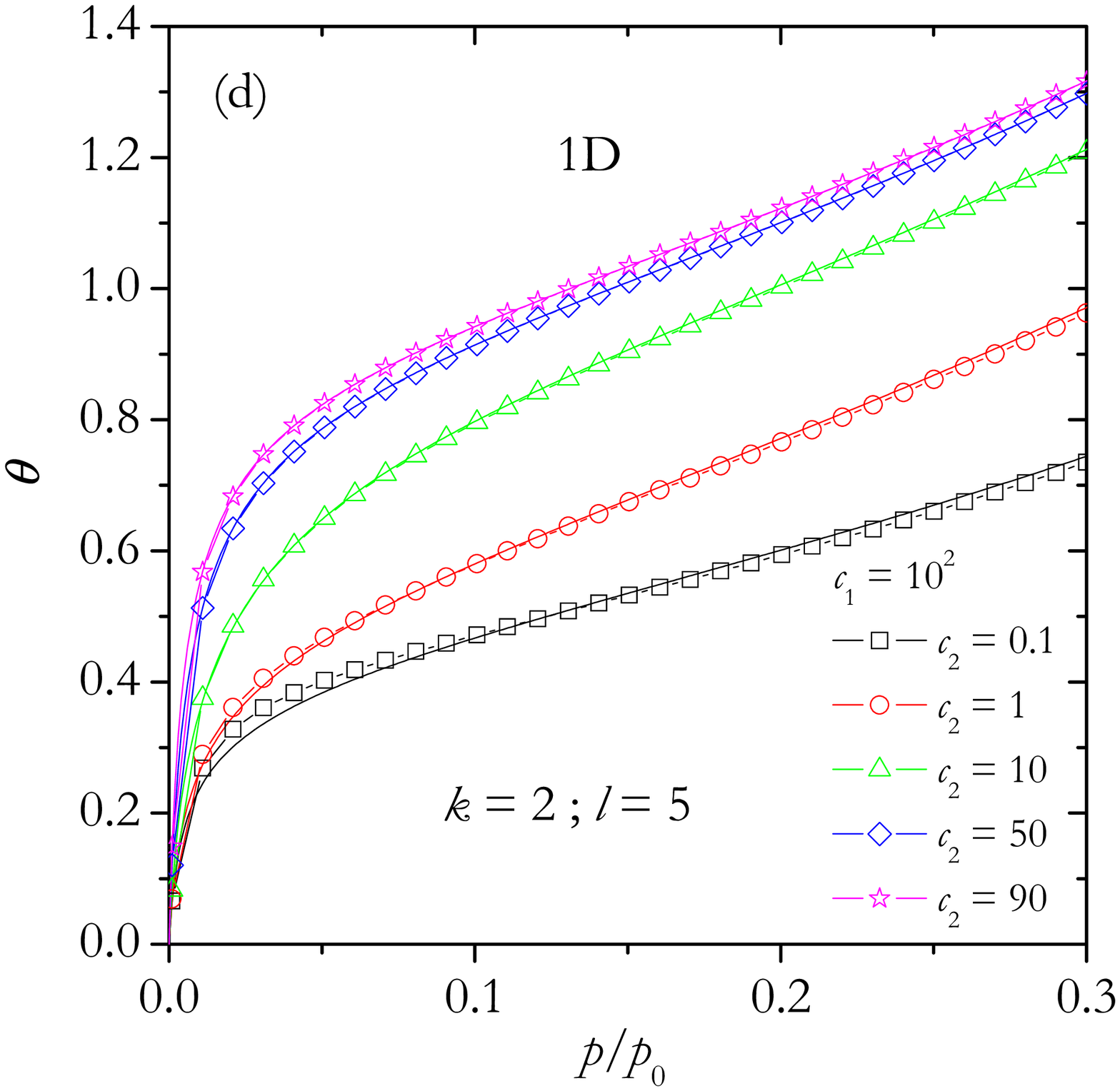}
}\caption{\label{figure3} Comparison between approximate analytic
and MC isotherms for dimers and different 1D heterogeneous
surfaces: (a) $l=2$, (b) $l=3$, (c) $l=4$ and (d) $l=5$. In all
cases we have used lattices of size $L=1200$ and $t=10^5$ MCSs.}
\end{figure}

For $l=1$, the adsorption energy of a dimer is $E= \varepsilon_1 +
\varepsilon_2$ for all configuration and, consequently,
eq~(\ref{isoh1Dk2}) is exact.  In general, for $l>1$ and $c_1 \neq
c_2$, this equation is approximate. Then, to determine the range
of validity of this equation, we compare the analytic isotherm
with MC results. Figure~\ref{figure3}a shows the dimer isotherm
for patches of size $l=2$, $c_1=10^2$ and different values of
$c_2$. As we can see, for $c_2 \geq 10$ the analytic isotherms
agree very well with the MC data. However, for smaller values of
$c_2$, the differences between theoretical and numerical data
begin to be significant. This happens because eq~(\ref{isoh1Dk2})
has been built assuming that the three different pairs of sites
are filled simultaneously and independently.  However, for $c_1
\gg c_2$, the real process occurs in 3 stages: $(i)$ the pairs of
sites (1,1) are covered; $(ii)$ the pairs (2,2) begin to be filled
and $(iii)$ the multilayer is formed. Note that in the first stage
all the pair of sites (1,2) and (2,1) are removed.  For this
regime, a better approximation can be obtained by a semisum of two
isotherms with $c_1$ y $c_2$.

When $l=3$, Figure~\ref{figure3}b, the agreement between the
analytic isotherms and the MC data is very good for all values of
$c_2$.  In this case, for $c_1 \gg c_2$ the first stage does not
eliminate all the pairs of sites (1,2) and (2,1), because each
dimer occupies only two sites in the strong patches. For this
reason, the range of validity of eq~(\ref{isoh1Dk2}) is wider than
in the previous case.  Now, if $l=4$ or $l=5$, we see in
Figures~\ref{figure3}c and d that the behaviors are similar to
those observed for $l=2$ or $l=3$, respectively.  In general, for
even $l$, the first stage eliminates almost completely the pairs
of sites (1,2) and (2,1), while this does not happen for odd $l$.
Finally, when $l \to \infty $, the fraction of pair (1,2) and
(2,1) goes to zero and eq~(\ref{isoh1Dk2}) is exact. This limit
corresponds to the called large patches topography (LPT), where
the surface is assumed to be a collection of homogeneous patches,
large enough to neglect border effects between neighbor patches
with different adsorption energies.

\begin{figure}[t]
\centering
\includegraphics[width=6.5cm,clip=true]{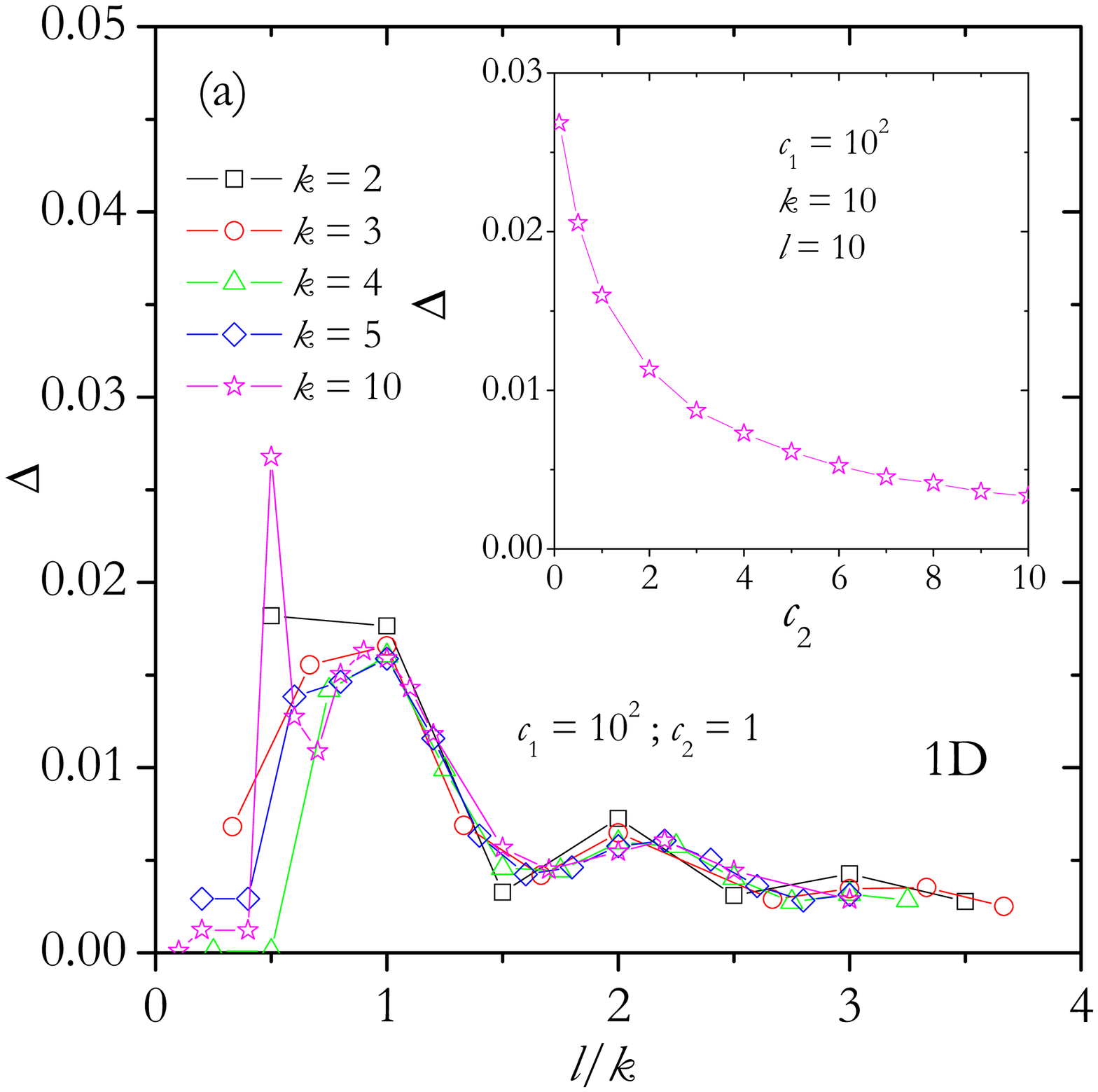}
\includegraphics[width=6.5cm,clip=true]{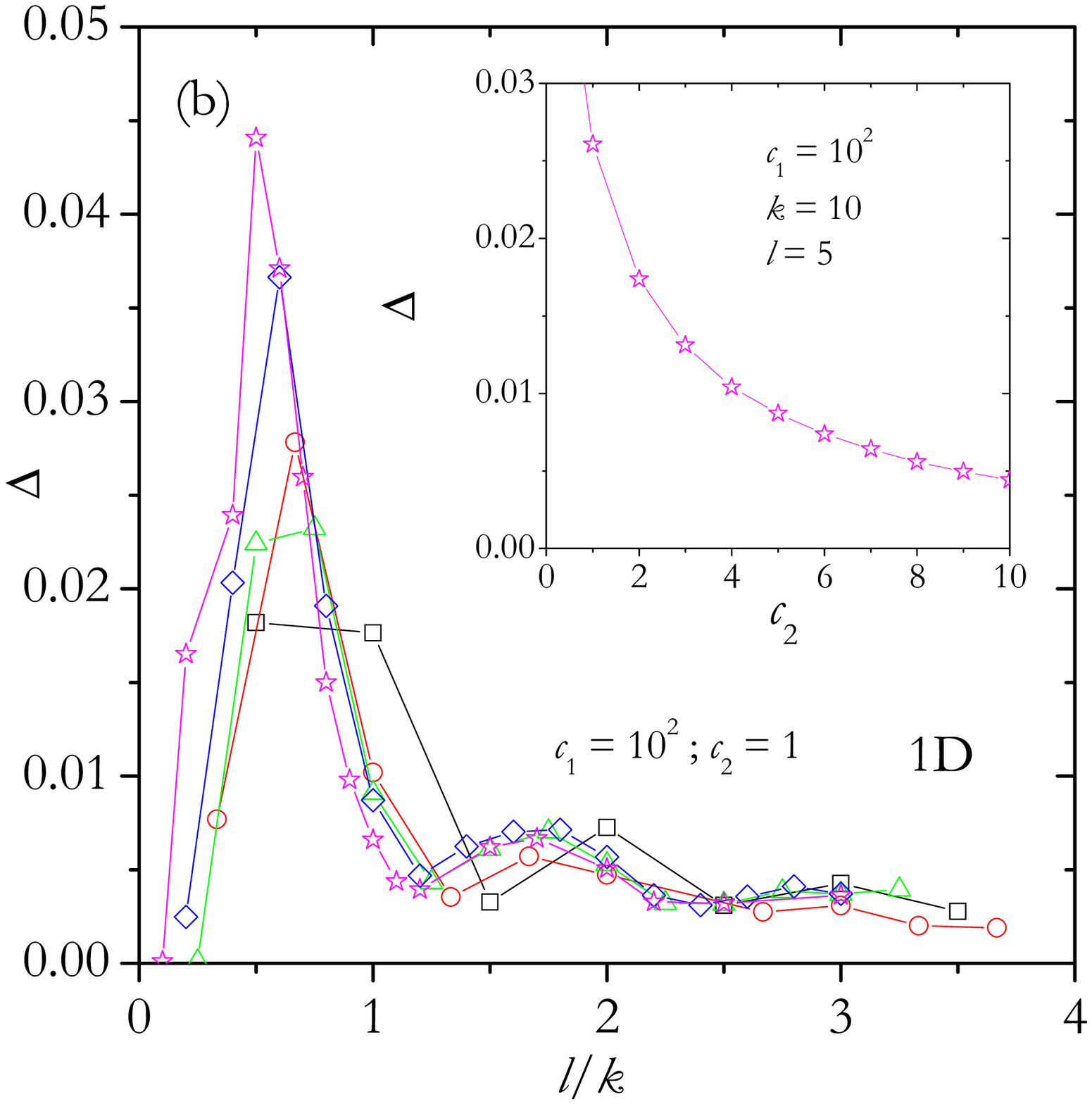}
\caption{\label{figure4} Function $\Delta$ vs $l/k$ for the
adsorption in a 1D heterogeneous surface with $c_1=10^2$, $c_2=1$
and different values of $k$ as indicated. (a) Function $\Delta$
calculated from eq~(\ref{isoh1D}). The inset shows the dependence
of $\Delta$ on $c_2$ for $k=10$ and $l=10$. b) Function $\Delta$
calculated from eq~(\ref{isoh1Daprox}). The inset shows the
dependence of $\Delta$ on $c_2$ for $k=10$ and $l=5$.}
\end{figure}

In general, if $k \gg l$ (with $k>1$), the multilayer adsorption
isotherm can be represented by a single homogeneous isotherm
\begin{equation}
\theta = \theta_\mathrm{loc} \left( \sqrt{c_1 c_2} \right).
\label{l1}
\end{equation}
On the other hand, for a LPT where $k \ll l$, the isotherm is
\begin{equation}
\theta = \frac{1}{2} \theta_\mathrm{loc} \left( c_1 \right)
+\frac{1}{2} \theta_\mathrm{loc} \left( c_2 \right). \label{LPT}
\end{equation}
The details of the topography are relevant only when $k \sim l$.
In this case, all terms in eq~(\ref{isoh1D}) are important.
Nevertheless, it is also interesting to obtain a simpler
expression of the multilayer isotherm given by
\begin{equation}
\theta = \left( \frac{l-1}{2l} \right) \theta_\mathrm{loc} \left(
c_1 \right) + \left( \frac{1}{l} \right) \theta_\mathrm{loc}
\left( \sqrt{c_1 c_2} \right) + \left( \frac{l-1}{2l} \right)
\theta_\mathrm{loc} \left( c_2 \right). \label{isoh1Daprox}
\end{equation}
Equation~(\ref{isoh1Daprox}) captures the extreme behaviors
eqs~(\ref{LPT}) and (\ref{l1}), and it approximates the MC
isotherm as well as eq~(\ref{isoh1D}).  To verify this statement,
we calculate the integral
\begin{equation}
\Delta=\int  \left|  \frac{\theta_{\mathrm{MC}} - \theta
}{\theta_{\mathrm{MC}}} \right| d(p/p_0), \label{delta}
\end{equation}
which allows to quantify the difference between the MC isotherm,
$\theta_{\mathrm{MC}}$, and the analytic isotherm, $\theta$, given
by either the eq~(\ref{isoh1D}) or the new approach
eq~(\ref{isoh1Daprox}).  For practical purposes, we have chosen a
range of relative pressure of $0-0.3$ to calculate the integral
eq~(\ref{delta}).  MC simulation were carried out for lattice
sizes of $L=20 k$ with a number of $t=10^5$ MCSs.

\begin{figure}[t]
\centerline{
\includegraphics[width=6.5cm,clip=true]{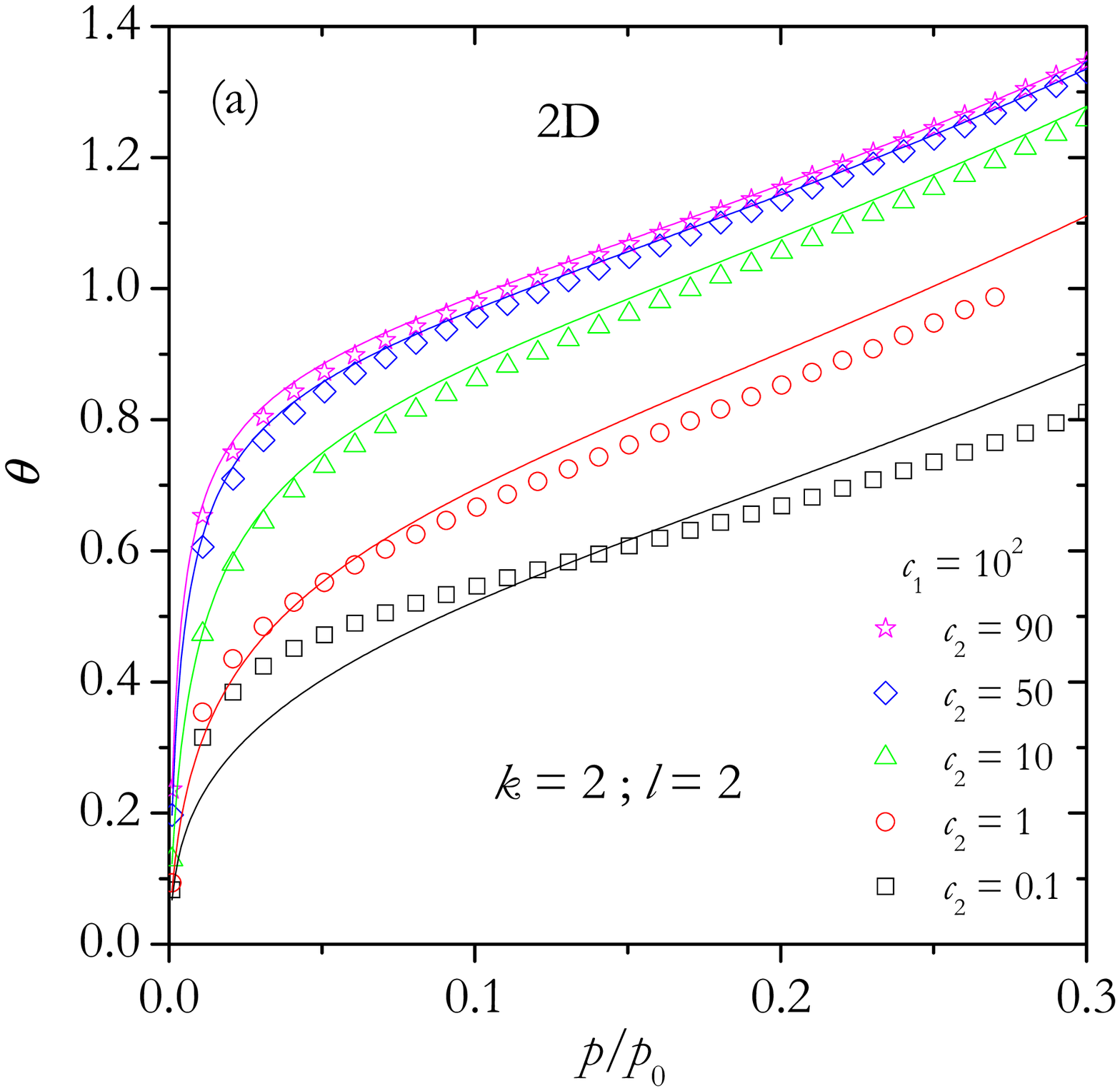}
\includegraphics[width=6.5cm,clip=true]{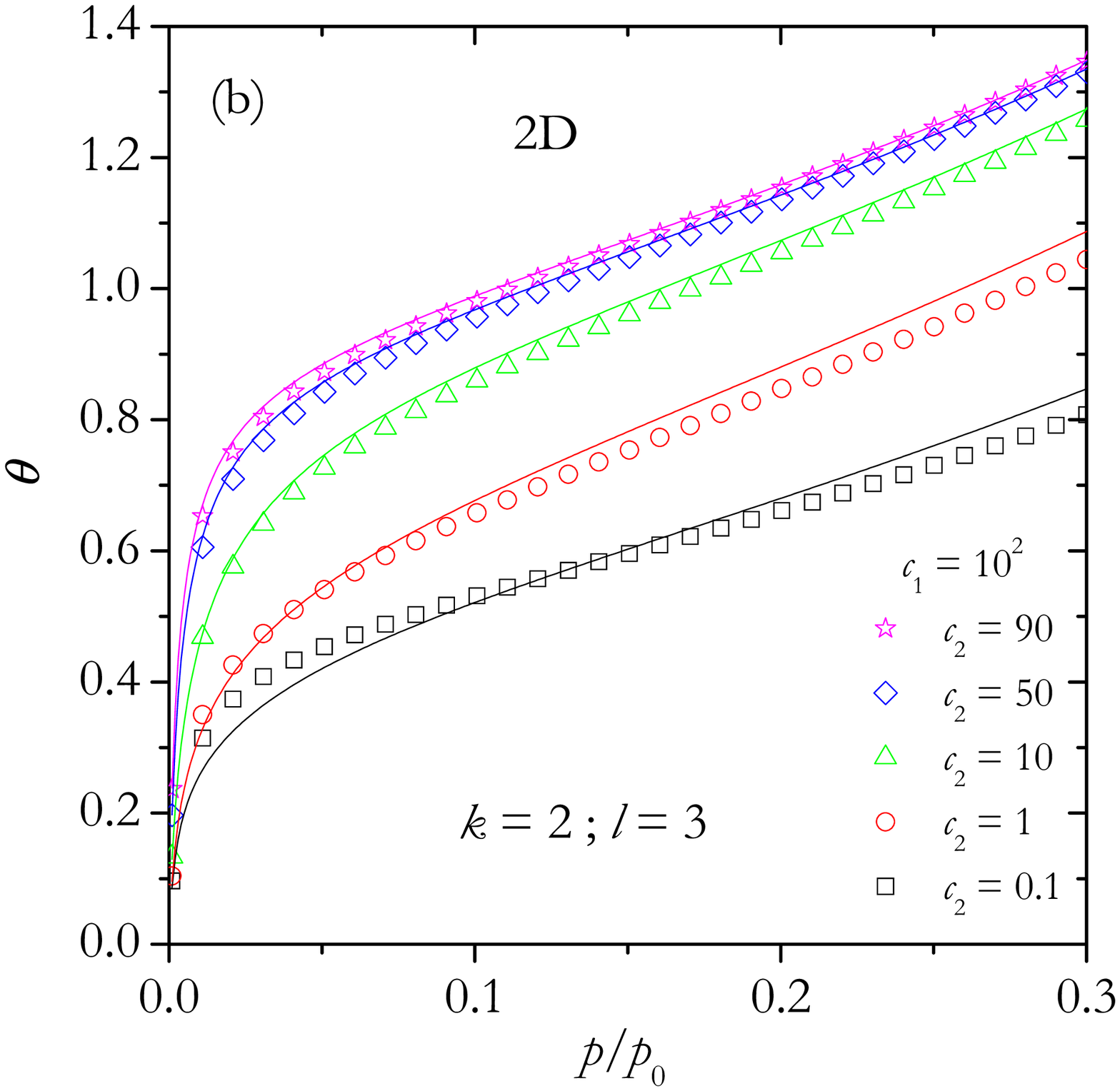}}
\centerline{
\includegraphics[width=6.5cm,clip=true]{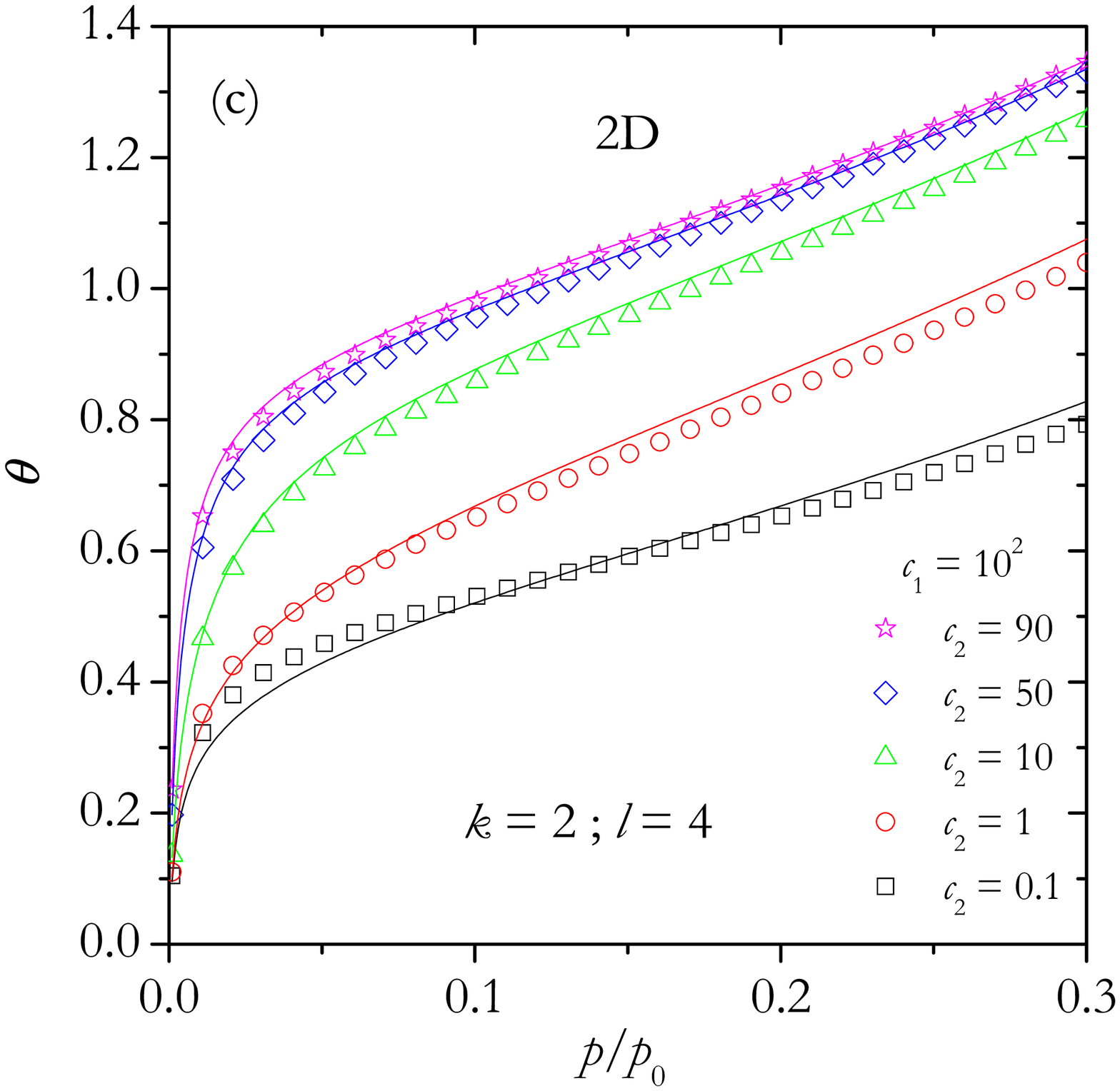}
\includegraphics[width=6.5cm,clip=true]{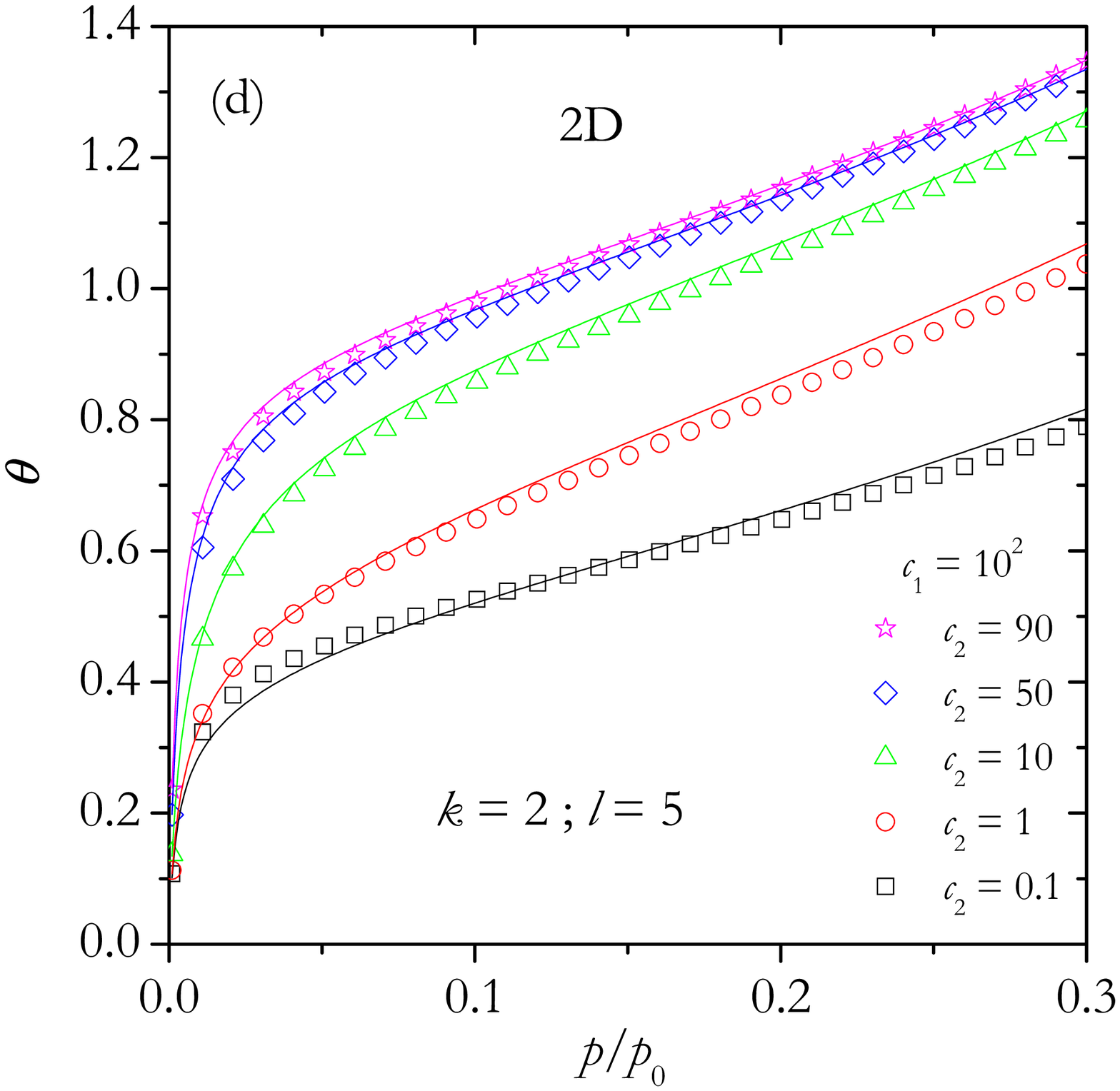}
} \caption{\label{figure5} Comparison between approximate analytic
and MC isotherms for dimers and different 2D heterogeneous
surface: (a) $l=2$, (b) $l=3$, (c) $l=4$ and (d) $l=5$. In all
cases, we have used lattices of size $L=240$ and $t=10^5$ MCSs.}
\end{figure}

Figure~\ref{figure4}a shows the function $\Delta$ calculated from
eq~(\ref{isoh1D}) for $c_1=10^2$, $c_2=1$ and different values of
$k$ and $l$.  As in the case of dimers, the difference between the
analytic and the MC isotherms increases when $l$ is approximately
a multiple of $k$. However, for large patches, i.e. $l > 3k$, this
difference becomes smaller.  The inset shows, for a particular
case ($k=10$ and $l=10$), how $\Delta$ diminishes as $c_2$ is
increased. On the other hand, in Figure~\ref{figure4}b we can see
the function $\Delta$ calculated from eq~(\ref{isoh1Daprox}). For
$k>2$, the first peak is higher than the one shown in
Figure~\ref{figure4}a and is located in a value of $l/k$ between
$0.5$ and $0.7$. Nevertheless, the oscillations attenuate quickly
as the parameter $l/k$ is increased. As before, but now for $k=10$
and $l=5$, the inset shows how $\Delta$ diminishes as $c_2$
increases.

Finally, the analysis of Figure~\ref{figure4} indicates that,
instead of eq~(\ref{isoh1D}) which has many terms,
eq~(\ref{isoh1Daprox}) can be used as a more simple expression to
approach the 1D multilayer adsorption isotherm.

\subsection{Adsorption on 2D Heterogeneous
Surfaces}

\begin{figure}[t]
\centering
\includegraphics[width=6.5cm,clip=true]{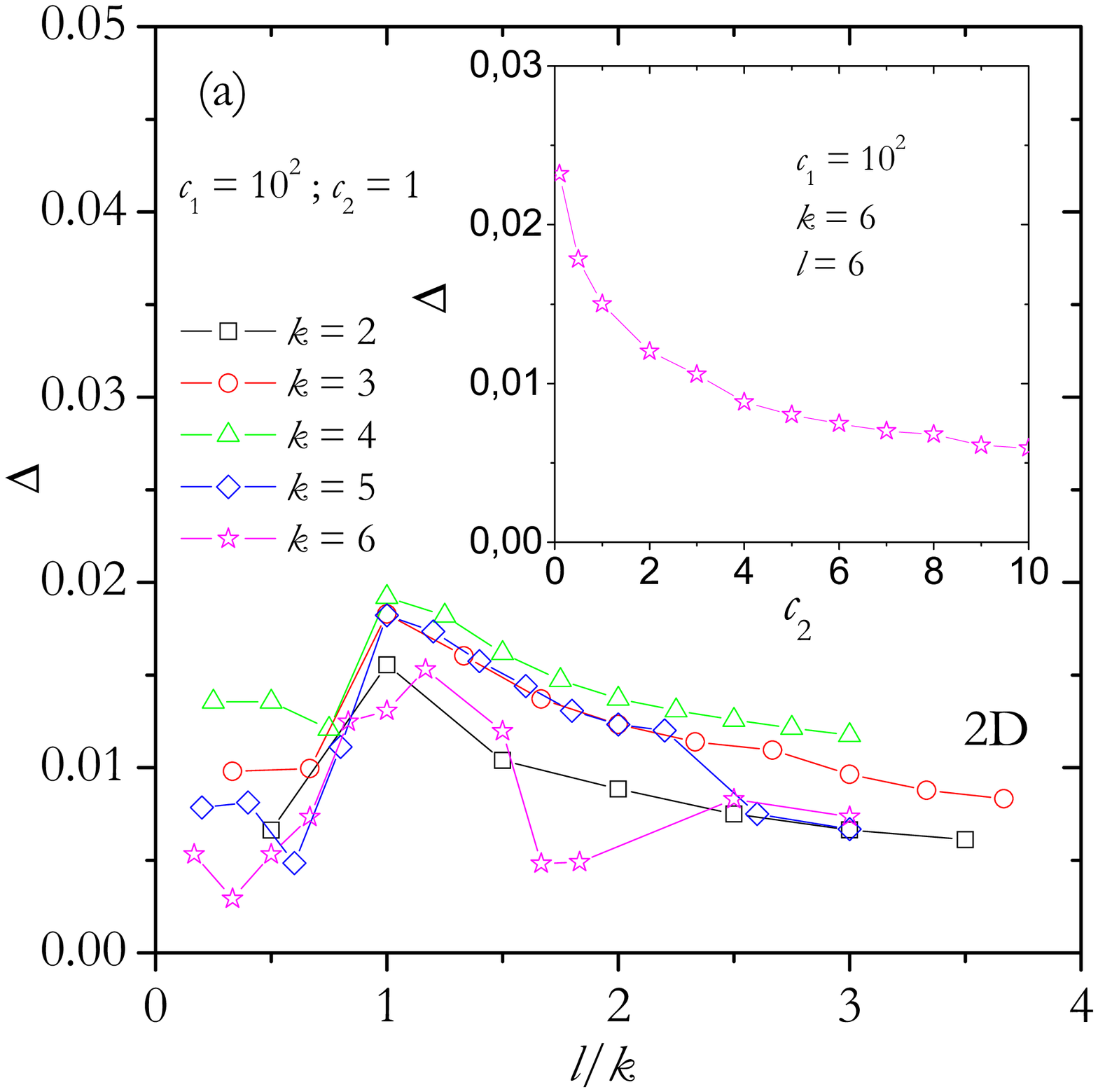}
\includegraphics[width=6.5cm,clip=true]{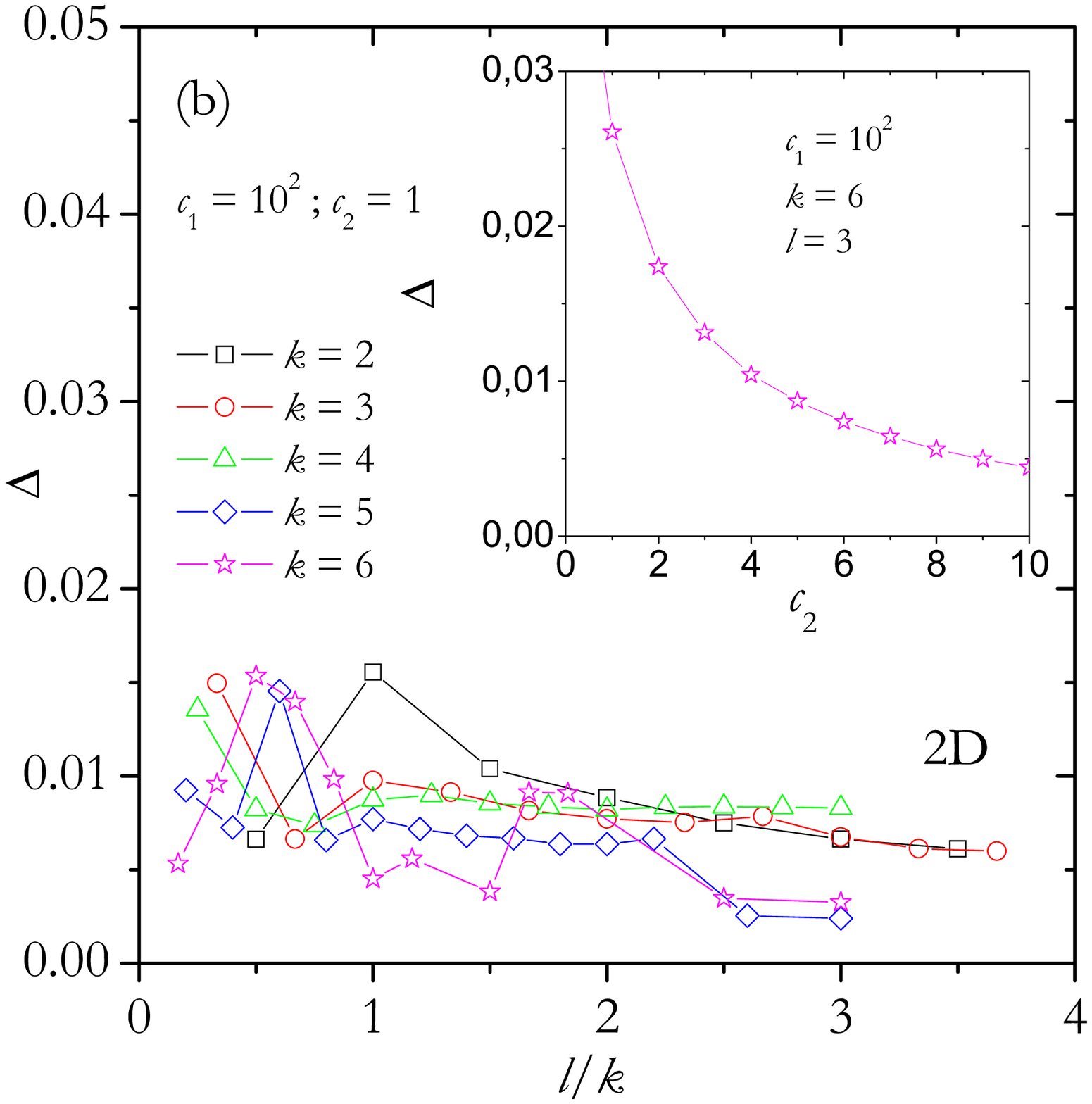}
\caption{\label{figure6} Function $\Delta$ vs $l/k$ for the
adsorption in a 2D heterogeneous surface with $c_1=10^2$, $c_2=1$
and different values of $k$ as indicated. (a) Function $\Delta$
calculated from eq~(\ref{isoh1D}). The inset shows the dependence
of $\Delta$ with $c_2$ for $k=6$ and $l=6$.  b) Function $\Delta$
calculated with eq~(\ref{isoh1Daprox}). The inset shows the
dependence of $\Delta$ with $c_2$ for $k=6$ and $l=3$.}
\end{figure}

As in the homogeneous case, we represent the 2D surface by a
square lattice with fully periodic boundary conditions.  Strong
and weak sites are spatially distributed in square patches of size
$l$ forming a chessboard. Now, the total number of configurations
of a single $k$-mer is $2M$.  However, as before, only $2l$ terms
are necessary to describe the multilayer isotherm. The explanation
is quite simple: the available energies of a $k$-mer that it is
forced to move in any direction of the lattice (row or column) are
the same that in 1D.  Then, eq~(\ref{isoh1D}) continues being
valid in 2D, where the local isotherm is given by
eqs~(\ref{pre2D}) and (\ref{cub}), with $\gamma = 4$.

We begin analyzing the multilayer isotherm for $k=2$ (for $k=1$,
eq~(\ref{isoh1Dk1}) continues being valid in 2D). Unfortunately,
it is not possible to write a simple analytic expression [as
eq~(\ref{isoh1Dk2})] in 2D. Nevertheless, the multilayer
adsorption isotherm for dimers has the same structure that
eq~(\ref{isoh1Dk2}). Namely, it is composed by three terms with
$c_1$, $c_2$ (both multiplied by $(l-1)/2l$) and $c=\sqrt{c_1
c_2}$ (multiplied by $1/l$).  Note that in 2D this function is
approximate for any value of the parameter $l$.

Figures~\ref{figure5}a, b, c and d show the multilayer isotherm
for $c_1=10^2$ and different values of $l$ and $c_2$.  The
behavior of these curves is very similar to the one observed in
1D, but the difference between analytic and MC adsorption
isotherms (for even and odd values of $l$) is no longer so
important. On the other hand, Figures~\ref{figure6}a and b show
the dependence of the function $\Delta$ on $l/k$, where $\Delta$
was calculated by using eq~(\ref{isoh1D}) and
eq~(\ref{isoh1Daprox}), respectively. As in the case of 2D
homogeneous surfaces, the analytic isotherm does not fit very well
the MC data for $k>6$. For this reason, Figures~\ref{figure6}a and
b show the function $\Delta$ up to $k=6$ only.

In addition, we have shown that just by using an expression of
three terms, eq~(\ref{isoh1Daprox}), we can approach very well the
multilayer isotherm in 1D and 2D for the adsorption on
heterogeneous surfaces.  In the next section, we will use this
approximation and MC simulations to study how the topography
affects the determination of monolayer volume predicted by the BET
equation.

\begin{figure}[t]
\centering
\includegraphics[width=6.5cm,clip=true]{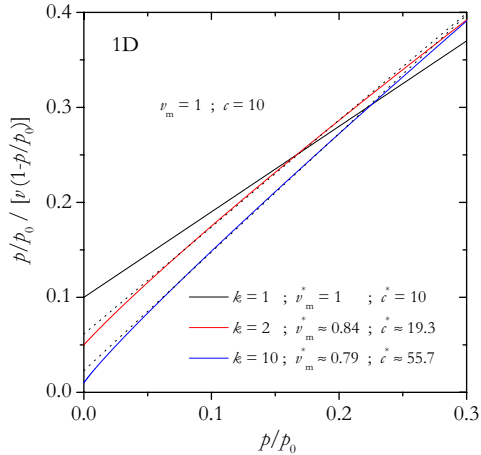}
\caption{\label{figure7} BET plots for 1D analytic isotherms of
$k$-mers with $v_\mathrm{m}=1$, $c=10$ and different values of $k$
as indicated. Dotted lines correspond to linear fits of the data
in the range $0.05$ to $0.25$.}
\end{figure}

\section{Monolayer Volume}

\begin{figure}[t]
\centering
\includegraphics[width=6.5cm,clip=true]{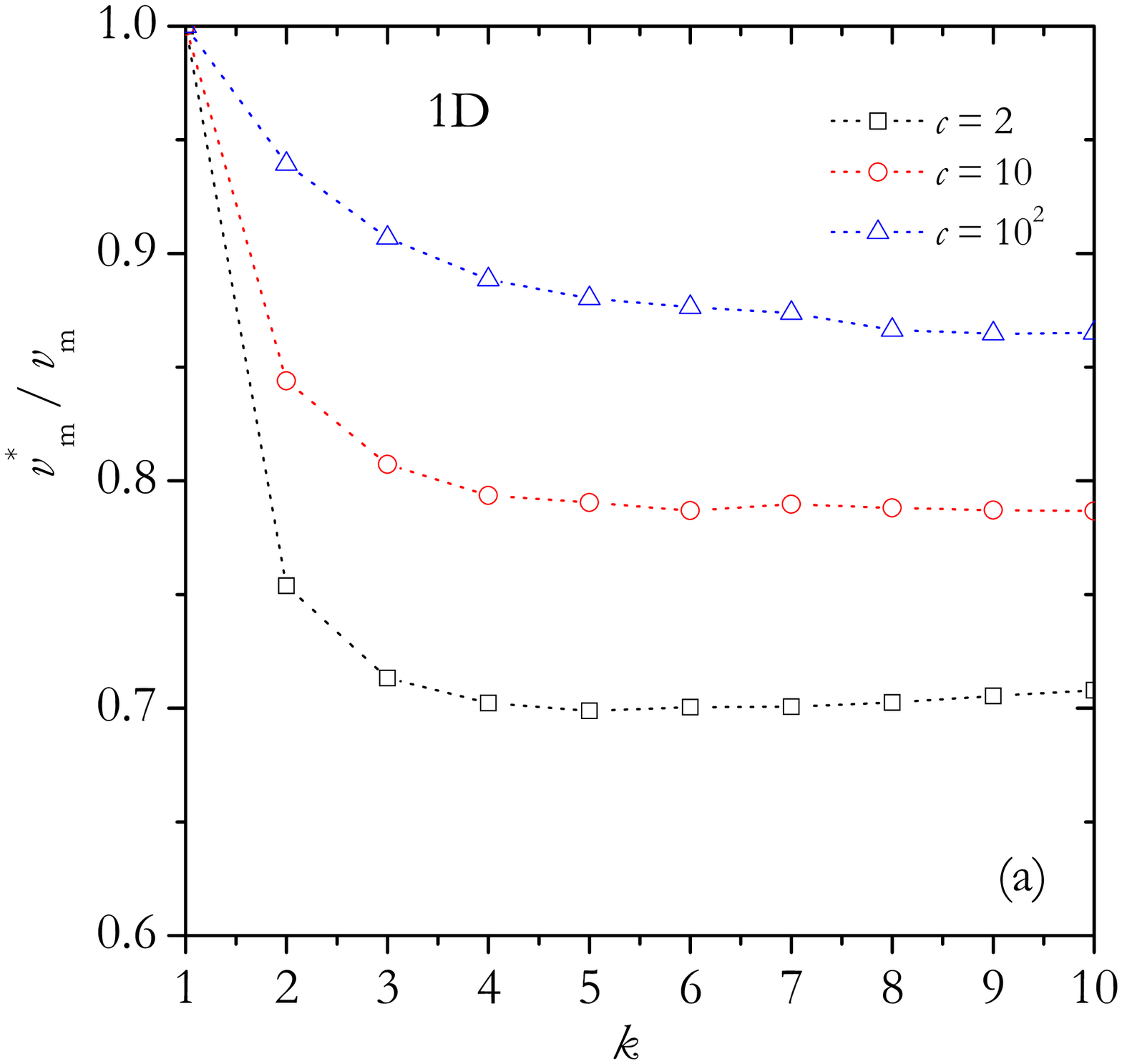}
\includegraphics[width=6.5cm,clip=true]{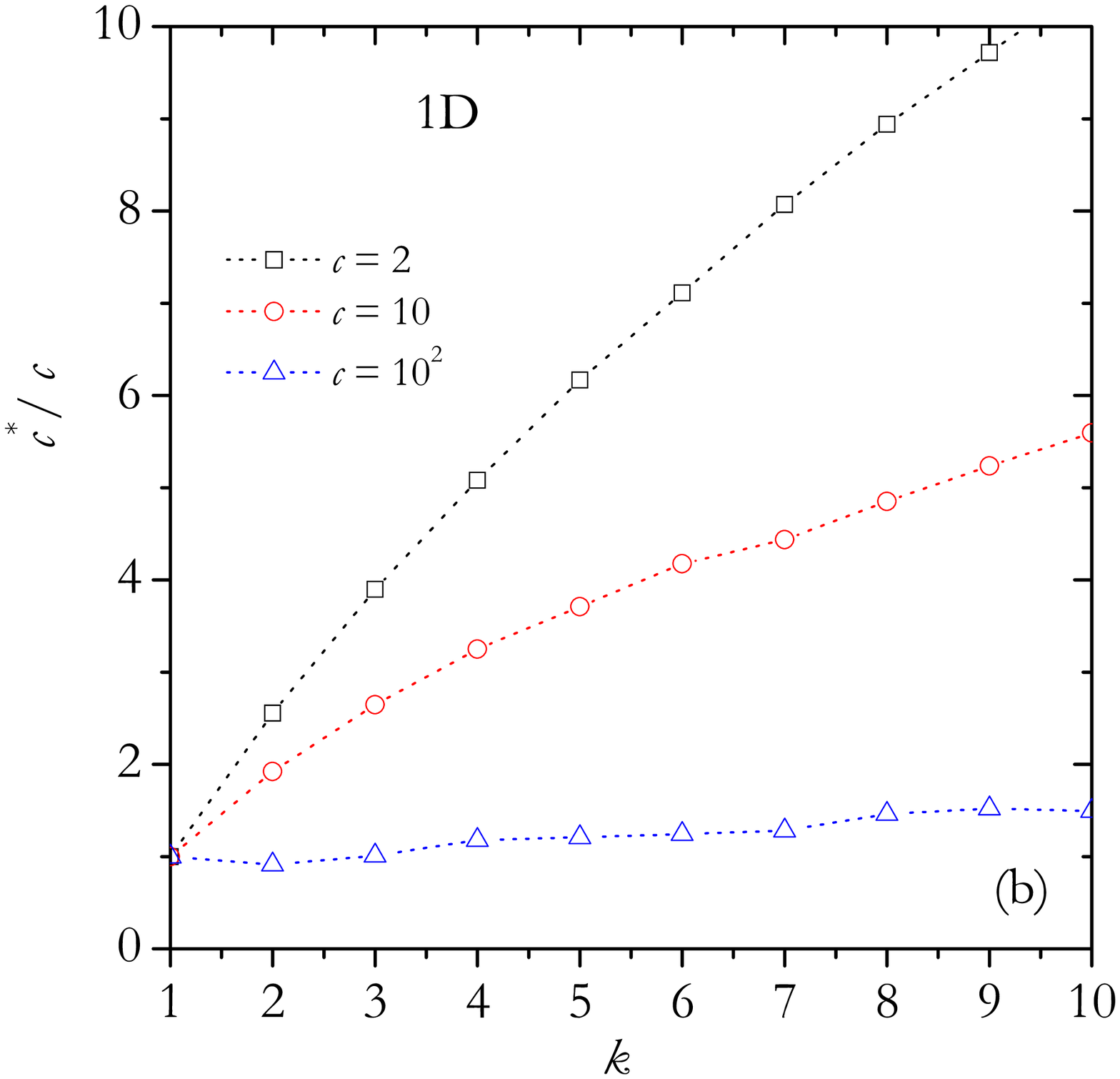}
\caption{\label{figure8} Results of the BET plots for the
adsorption in 1D homogeneous surfaces.  Dependence on $k$ of the
fractions (a) $v_\mathrm{m}^* / v_\mathrm{m}$ and (b) $c^* / c$,
for three different values of $c$ as indicated.}
\end{figure}

In this section, we carry out numerical experiments to determine,
in different adsorption situations, how much the  value of the
monolayer volume predicted by the BET equation differs from its
real value, $v_\mathrm{m}$. With this purpose, analytic and MC
isotherms were analyzed as experimental data. In this way, we have
determined how adsorbate size, energetic heterogeneity and surface
topography affect the standard determination of the monolayer
volume.

In a typical experiment of adsorption, the adsorbed volume of the
gas, $v$, is measured at different pressures and at a given fixed
temperature. In terms of this quantity, the total coverage is
$\theta=v/v_\mathrm{m}$. Analyzing an isotherm with the BET
equation, it is possible to estimate the monolayer volume.  We
rewrite the eq~(\ref{BET}) as
\begin{equation}
\frac{p/p_0}{v \left( 1- p/p_0 \right)}=\frac{1}{c
v_\mathrm{m}}+\frac{\left( c-1 \right)}{c v_\mathrm{m}} p/p_0 .
\label{BETlin}
\end{equation}
This equation is a linear function of $p/p_0$.  If we denote with
$a$ and $b$, the $y$-intercept and the slope of this straight
line, respectively, we obtain
\begin{equation}
v_\mathrm{m}^*=\frac{1}{a+b} \label{vas}
\end{equation}
and
\begin{equation}
c^*=\frac{b}{a}+1 .\label{cas}
\end{equation}
The asterisk has been added in order to indicate that the
quantities given by eqs~(\ref{vas}) and (\ref{cas}) correspond to
the prediction of the BET theory. Then, by means of a plot (the
so-called BET plot) of the experimental data of $\frac{p/p_0}{v
\left( 1- p/p_0 \right)}$ vs $p/p_0$, we can obtain an estimate of
the monolayer volume and the parameter $c$. Nevertheless, in the
experiments it is commonly found that there are deviations from
linearity in the BET plot. In many cases, the linear range extends
from a relative pressure of $0.05$ to $0.35$, although there are
cases where the range is shorter. \cite{Gregg1991}

Although the BET plot is a very simple and popular protocol, the
value of the monolayer volume obtained in this way can differ from
its real value. As mentioned in the introduction, in an
interesting numerical experiment, \cite{Walker1948} Walker and
Zettlemoyer analyzed a BET plot of an analytic isotherm composed
by two BET-like contributions (a isotherm similar to
eq~(\ref{isoh1Dk1}) for a LPT), each one with different values of
$v_\mathrm{m}$ and $c$. The authors concluded that the application
of the conventional BET equation to this heterogeneous isotherm
may lead to an underestimate of the true monolayer volume, with a
$c$ lying between the values for the two type of sites. Later,
Cort\'es and Araya \cite{Cortes1987} have obtained a similar
result by averaging the BET equation with a Gaussian distribution
of adsorption energy.  More recently Nikitas, \cite{Nikitas1996}
has arrived to similar conclusions by considering both, surface
heterogeneity and polyatomic character of the adsorbate.

\begin{figure}[t]
\centering
\includegraphics[width=6.5cm,clip=true]{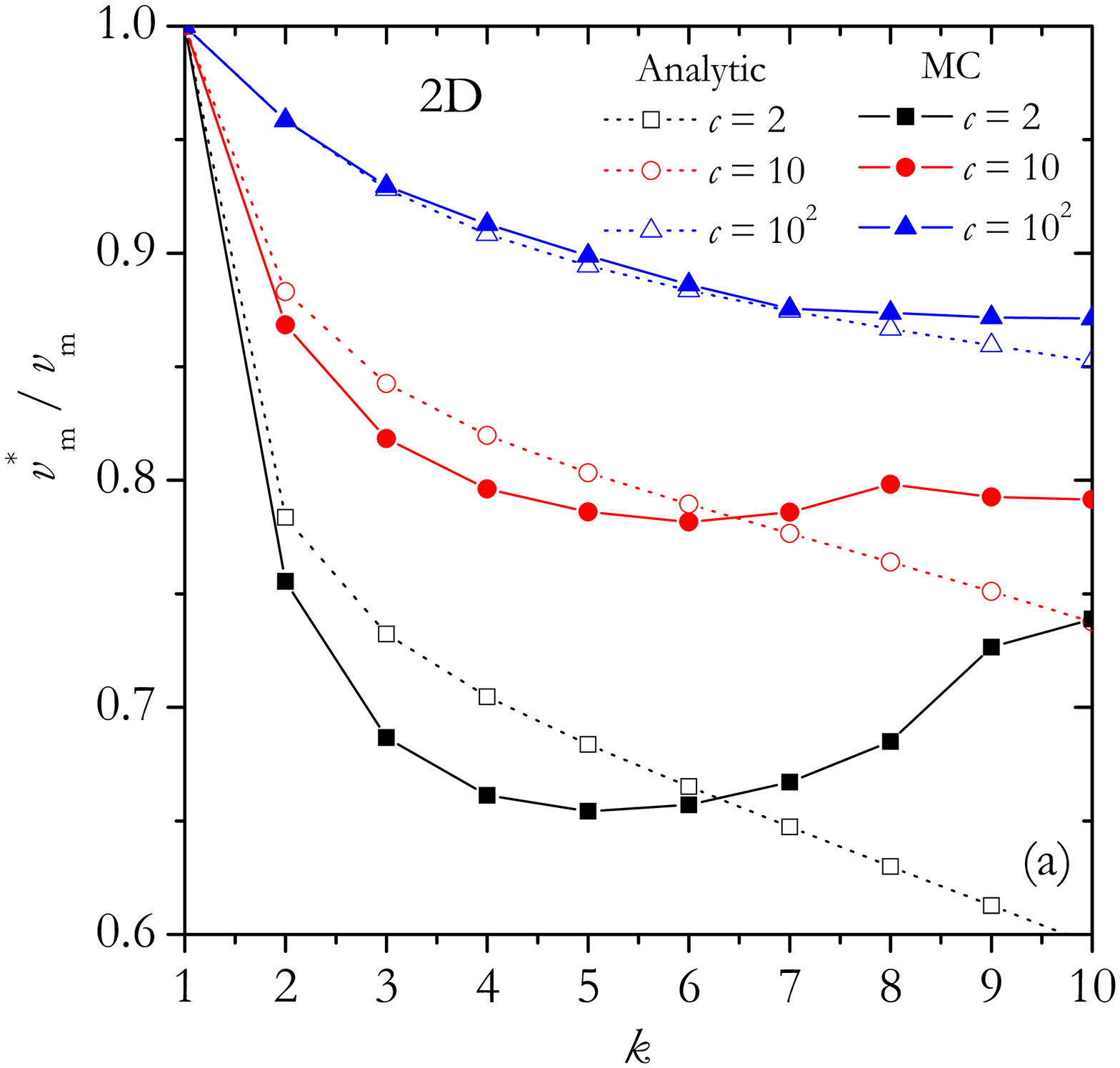}
\includegraphics[width=6.5cm,clip=true]{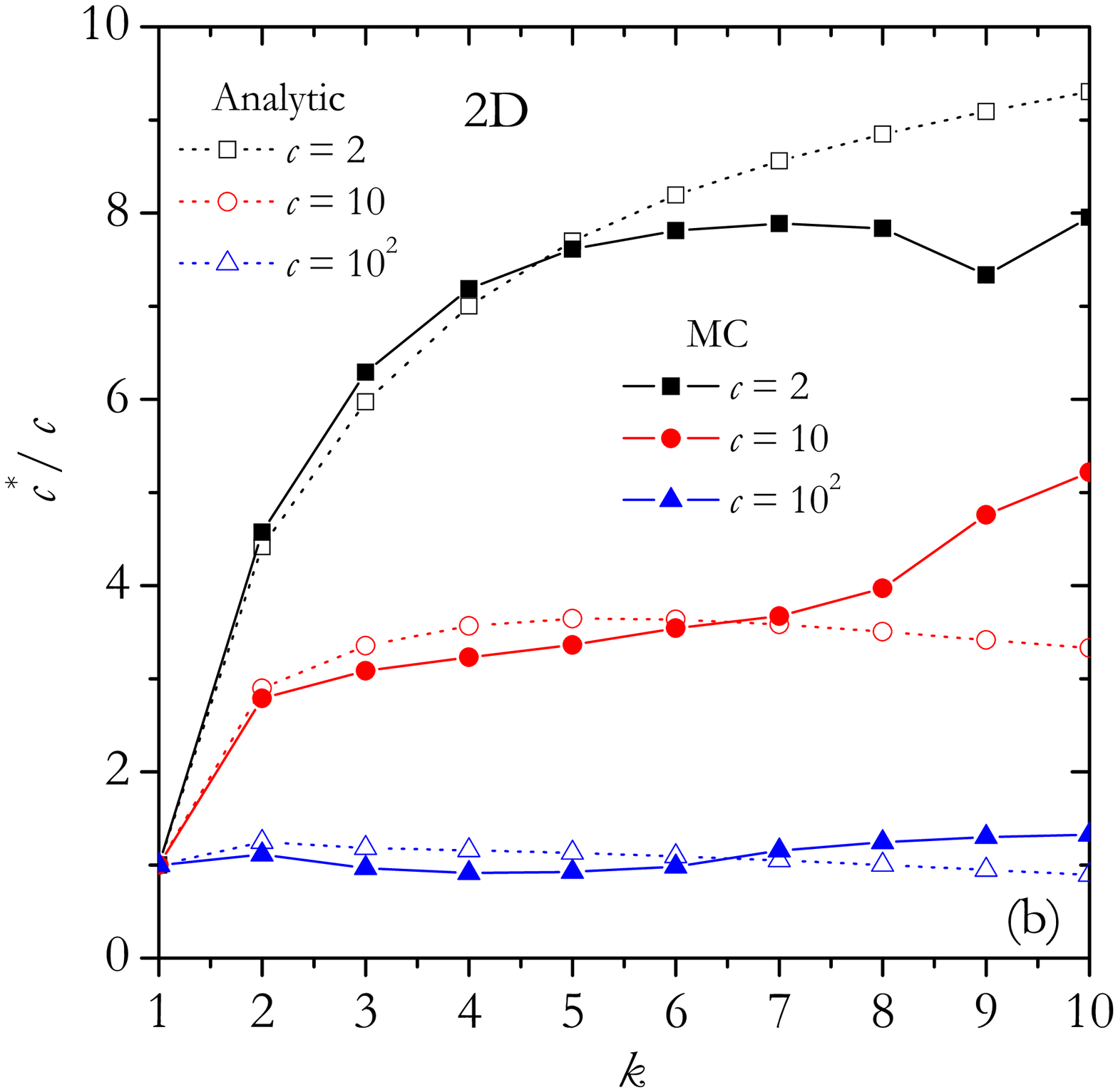}
\caption{\label{figure9} Results of the BET plots for the
adsorption in 2D homogeneous surfaces.  Dependence on $k$ of the
fractions (a) $v_\mathrm{m}^* / v_\mathrm{m}$ and (b) $c^* / c$,
for three different values of $c$ as indicated.}
\end{figure}

In the following, we will show that even for adsorption over
homogeneous surfaces, the polyatomic character of the adsorbate
affects significantly the predictions of a BET plot. Next, in
Section 5.2, by considering bivariate surfaces, we will study the
combined effect of energetic heterogeneity and multisite
occupancy.

\subsection{Homogeneous Surfaces}

We begin analyzing the BET plots of the multilayer adsorption of
$k$-mers over homogeneous surfaces, given by eqs~(\ref{cub}) and
(\ref{pre2D}) and MC data. Although in each particular case it is
possible to find an optimum range of relative pressures, for
practical purposes, we have chosen to set this range from $0.05$
to $0.25$. Nevertheless, by choosing other ranges (for example,
between $0.05$ and $0.35$) we obtain similar results.

In Figure~\ref{figure7} we show the BET plot for 1D analytic
isotherms with $c=10$ and $k=1$, 2 and 10.  Note the deviations
from linearity in the isotherms for $k=2$ and 10, which are
concave to the pressure axis.  The same behavior is observed in
experimental isotherms and it is attributed to the existence of
surface heterogeneities. \cite{Walker1948} However, as we see in
the example shown in Figure~\ref{figure7}, these deviations also
appear for the multilayer adsorption with multisite occupancy on a
homogeneous surface.

On the other hand, as indicated in Figure~\ref{figure7}, the
obtained value of $v_\mathrm{m}^*$ for $k>1$ is smaller than the
real one (we set $v_\mathrm{m}=1$), while the opposite effect is
observed in the estimate of the parameter $c$.
Figures~\ref{figure8}a and b show the dependence of these
quantities on $k$ for different values of $c$.  In all cases, we
obtain $v_\mathrm{m}^* \leq v_\mathrm{m}$ and $c^* \geq c$, but
the differences between the BET predictions and the real values
are smaller with increasing $c$.

\begin{figure}[t]
\centering
\includegraphics[width=6.5cm,clip=true]{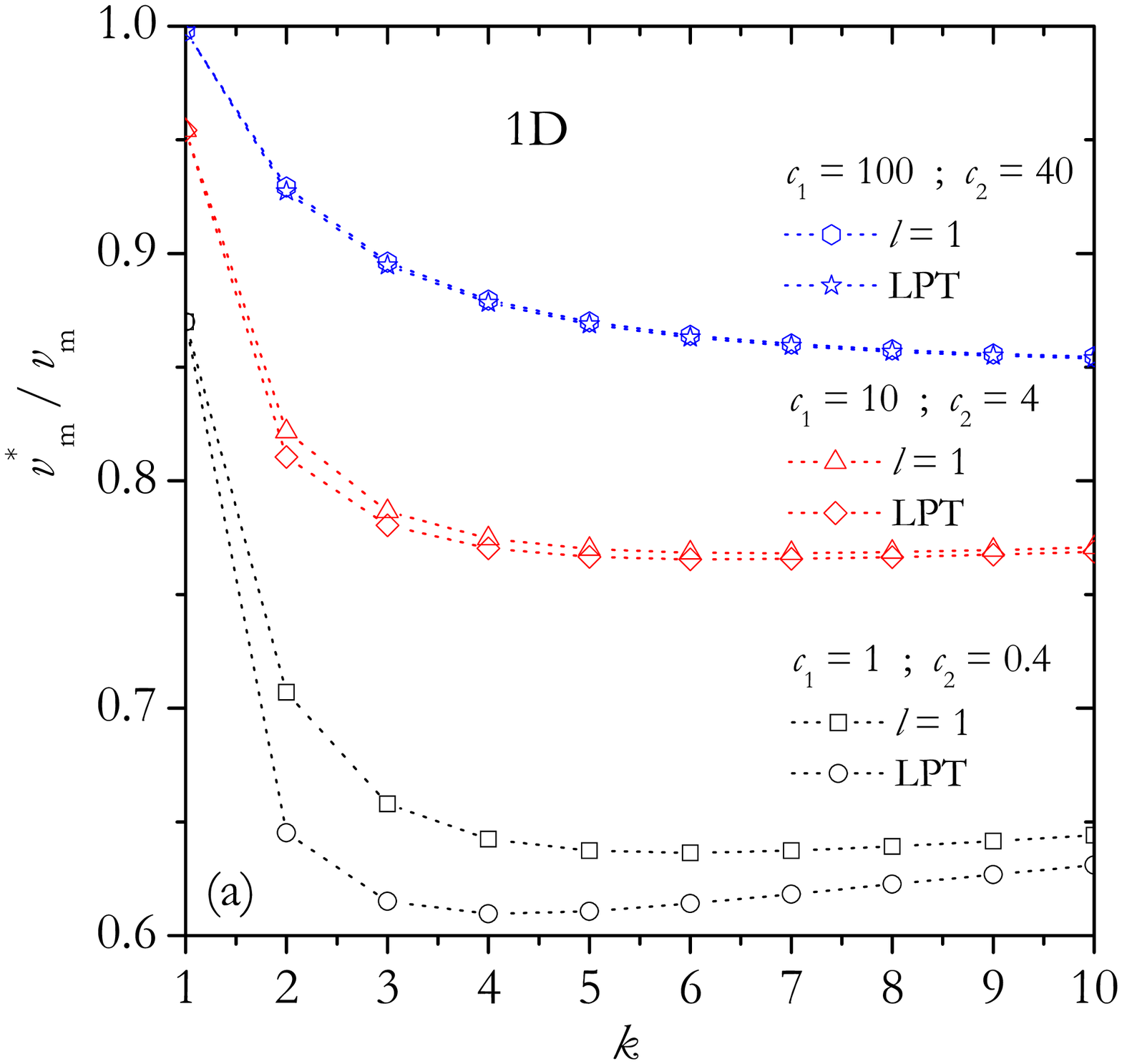}
\includegraphics[width=6.5cm,clip=true]{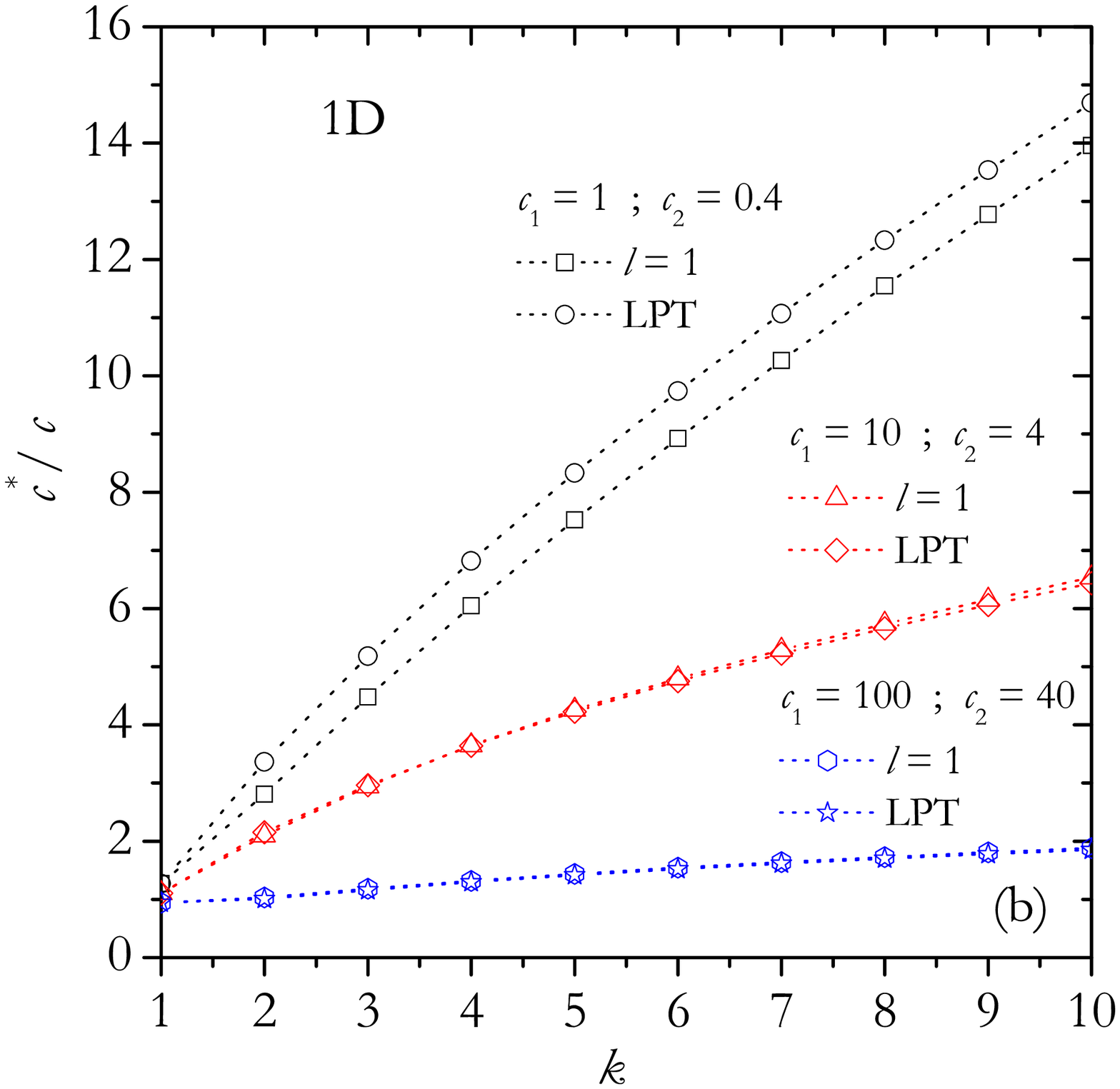}
\caption{\label{figure10} Results of the BET plots for the
adsorption in 1D heterogeneous surfaces with $l=1$ and LPT.
Dependence on $k$ of the fractions (a) $v_\mathrm{m}^* /
v_\mathrm{m}$ and (b) $c^* / c$, for three different values of
$c_1$ and $c_2$ (being $c_1 / c_2 =2.5$) as indicated.}
\end{figure}

Similar results have been obtained in 2D: the BET plots of both
analytic and MC isotherms show the same curvature as found in 1D.
Figures~\ref{figure9}a and b show the results of these 2D BET
plots. As we can see, the differences between analytic and MC
isotherms are significant for small values of $c$.  However,
always $v_\mathrm{m}^* \leq v_\mathrm{m}$ and $c^* \geq c$ for
$k>1$.  As in the 1D case, the monolayer volume predicted by BET
is approximately 10-30 per cent smaller than the real value.

\subsection{Heterogeneous Surfaces}

In previous work, \cite{Walker1948,Cortes1987} it has been
determined that, as heterogeneous adsorption isotherms of monomers
are analyzed, the monolayer volume obtained from a BET plot is
smaller than the real value. Since in this case $k=1$, the surface
topography does not affect the obtained results. In this section,
we study the dependence of the monolayer volume on both, adsorbate
size and surface topography.  In particular, we analyze analytic
and MC adsorption isotherms of $k$-mers over bivariate surfaces
with $l=1$ and LPT.

\begin{figure}[t]
\centering
\includegraphics[width=6.5cm,clip=true]{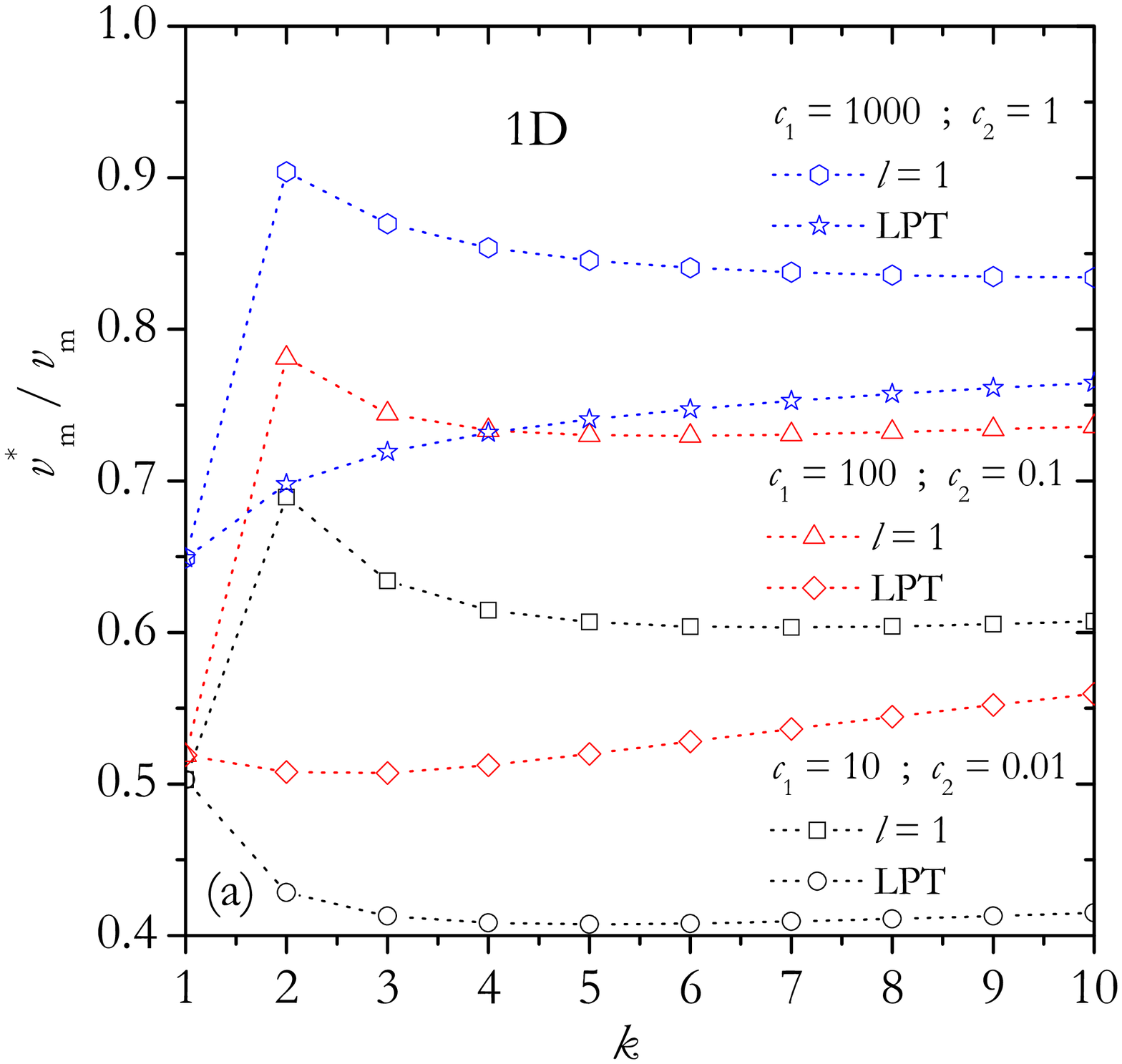}
\includegraphics[width=6.5cm,clip=true]{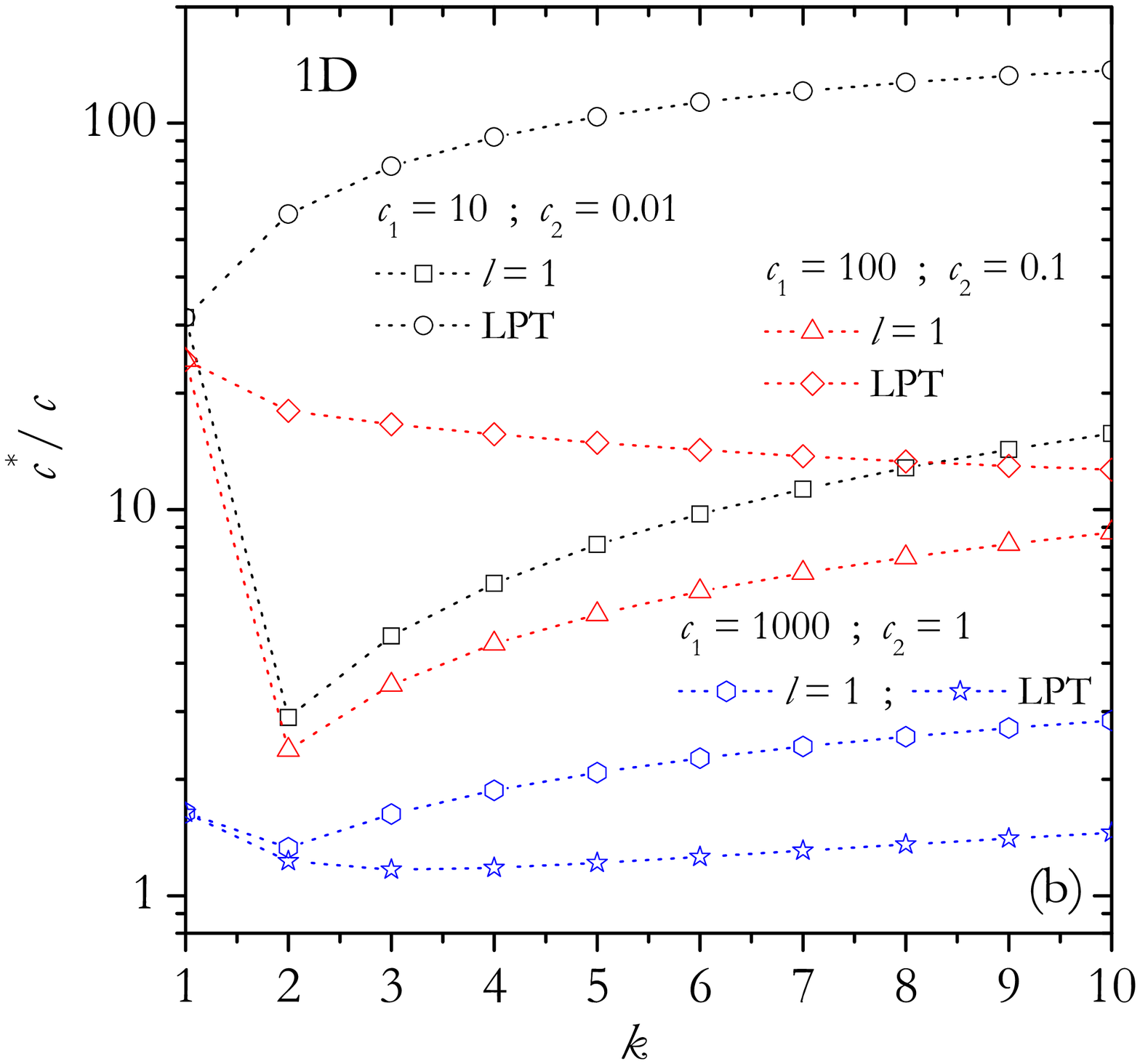}
\caption{\label{figure11} Results of the BET plots for the
adsorption in 1D heterogeneous surfaces with $l=1$ and LPT.
Dependence on $k$ of the fractions (a) $v_\mathrm{m}^* /
v_\mathrm{m}$ and (b) $c^* / c$, for three different values of
$c_1$ and $c_2$ (being $c_1 / c_2 =10^3$) as indicated.}
\end{figure}

Figures~\ref{figure10}a and b show the results of the 1D BET plots
for three different values of $c_1$ and $c_2$, being $c_1 / c_2
=2.5$. In this case, only analytic isotherms were studied because
they are exact for LPT and the agreement with MC data is seen to
be remarkably good for $l=1$ [except for odd values of $l$
($l>1$), as was previously mentioned]. In all cases we have used
$c=\sqrt{c_1 c_2}$ as the reference parameter. As we can see, the
curves show that there is not a significant difference between
both topographies. Only when the quotient between $c_1$ and $c_2$
is increased, the space distribution of the adsorption energies
over the solid surface begins to be important.  This is shown in
Figures~\ref{figure11}a and b, where $c_1 / c_2 =10^3$. The
results of the BET plots for $l=1$ and LPT are very different. For
$c_1=10$ and $c_2=0.01$, the deviations due to molecule size $k$
are increased in LPT, i. e. the monolayer volume and the parameter
$c$ obtained from a BET plot are, respectively, smaller and larger
than the real values (or the reference value). However, most of
the curves show a compensation effect which is larger for $l=1$,
and for $c_1=10^3$ and $c_2=1$.

\begin{figure}[t]
\centering
\includegraphics[width=6.5cm,clip=true]{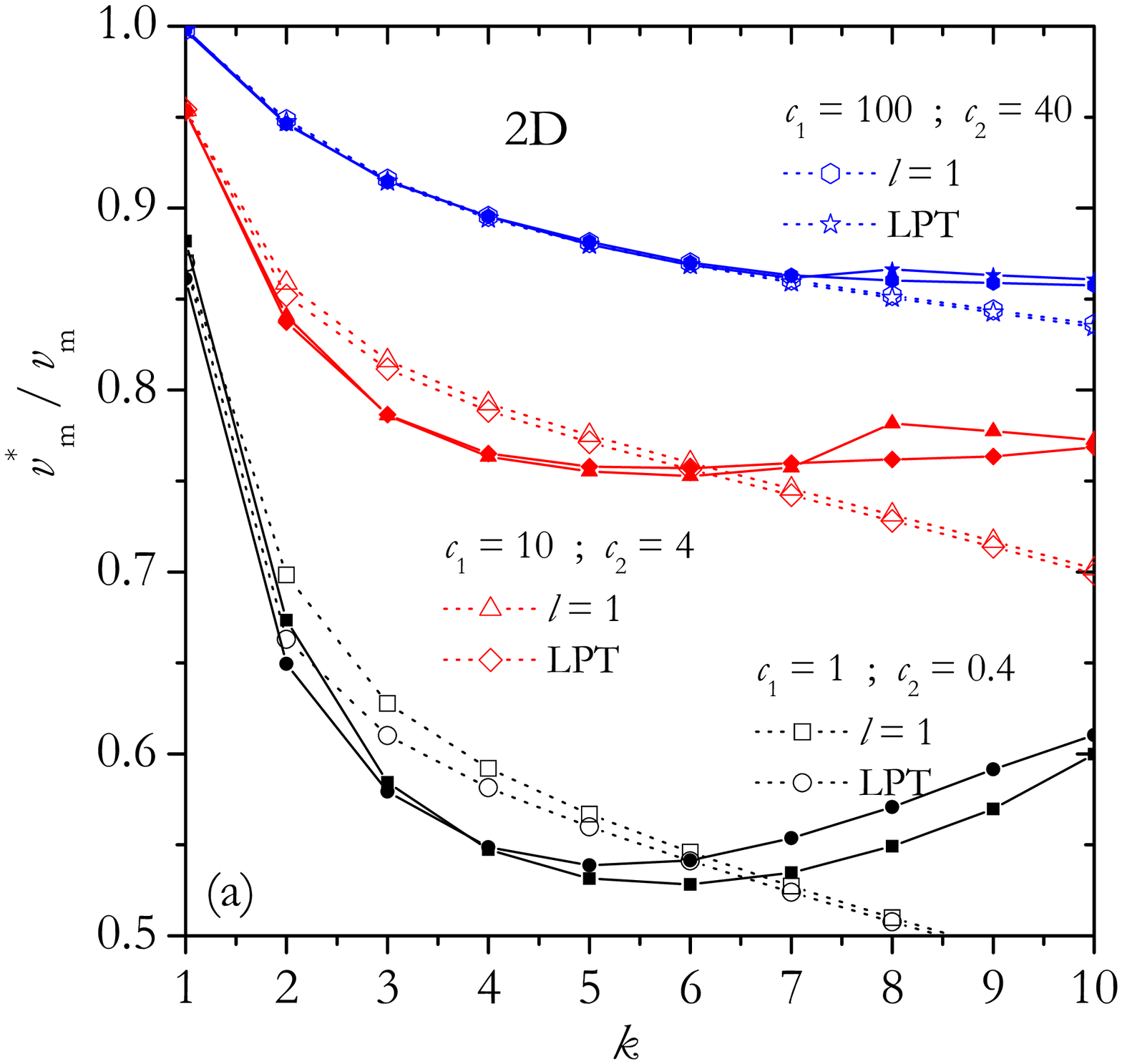}
\includegraphics[width=6.5cm,clip=true]{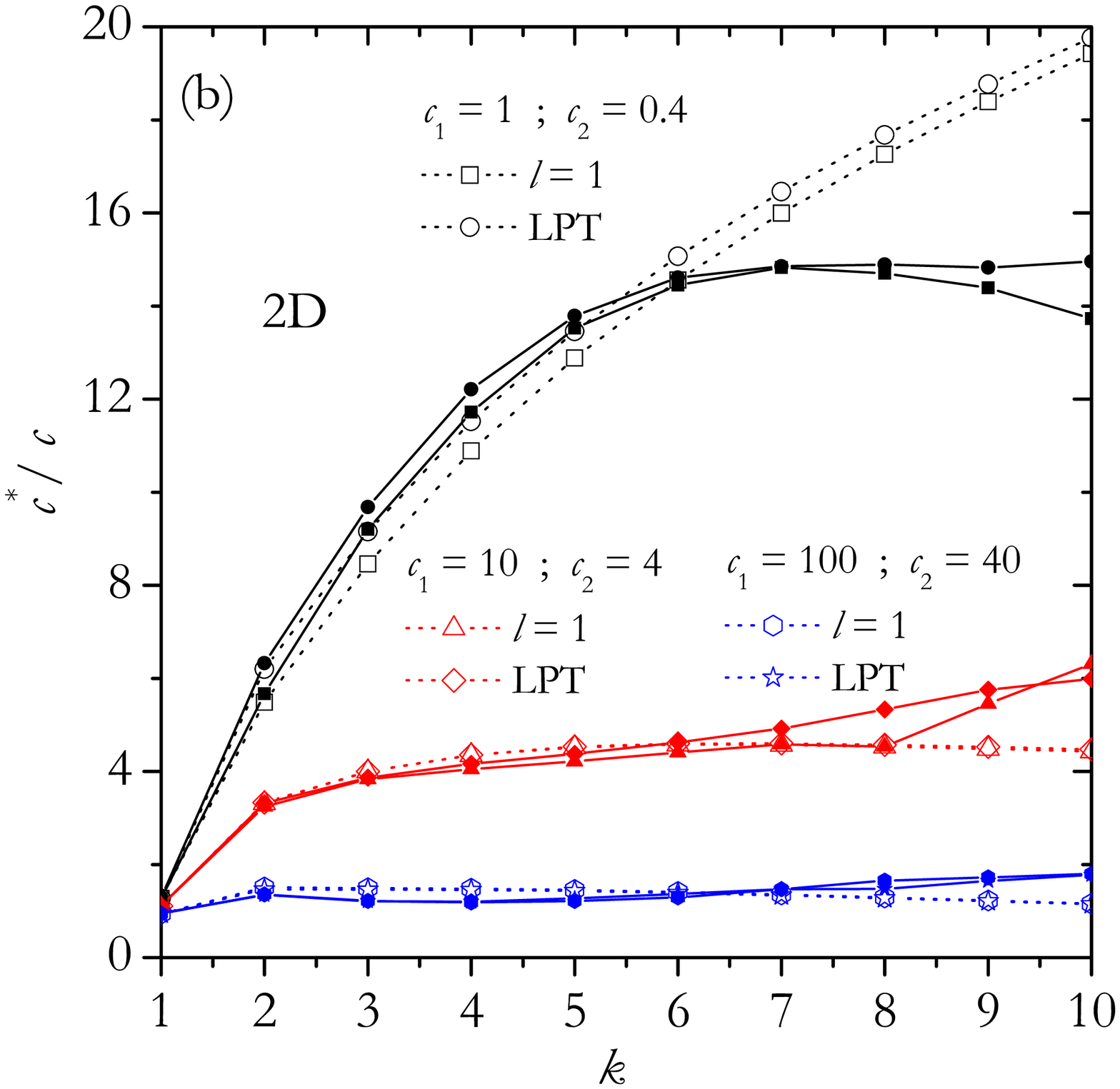}
\caption{\label{figure12} Results of the BET plots for the
adsorption in 2D heterogeneous surfaces with $l=1$ and LPT. Open
(full) symbols show results obtained from the analytic (MC)
isotherms. Dependence on $k$ of the fractions (a) $v_\mathrm{m}^*
/ v_\mathrm{m}$ and (b) $c^* / c$, for three different values of
$c_1$ and $c_2$ (being $c_1 / c_2 =2.5$) as indicated.}
\end{figure}

\begin{figure}[t]
\centering
\includegraphics[width=6.5cm,clip=true]{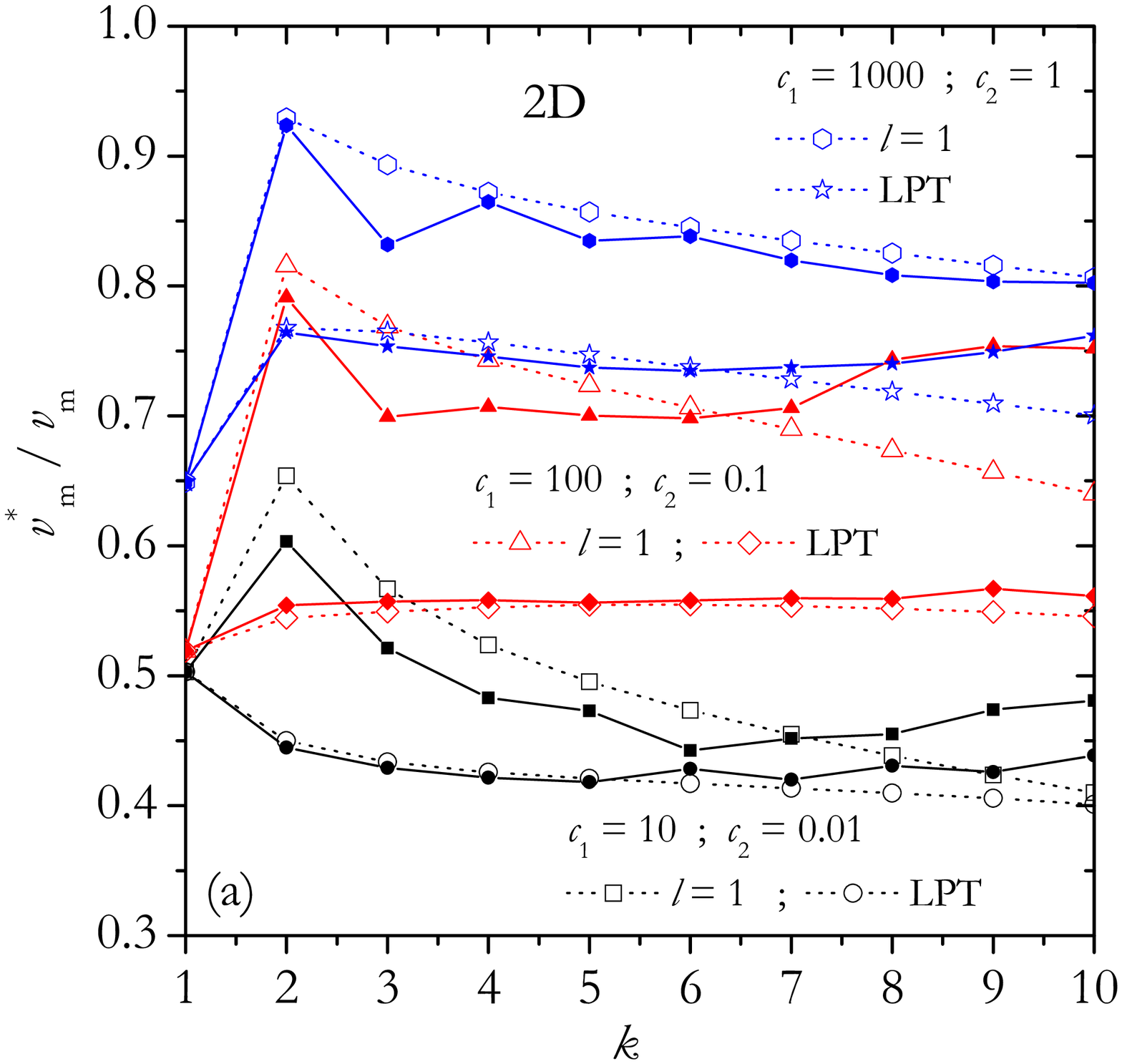}
\includegraphics[width=6.5cm,clip=true]{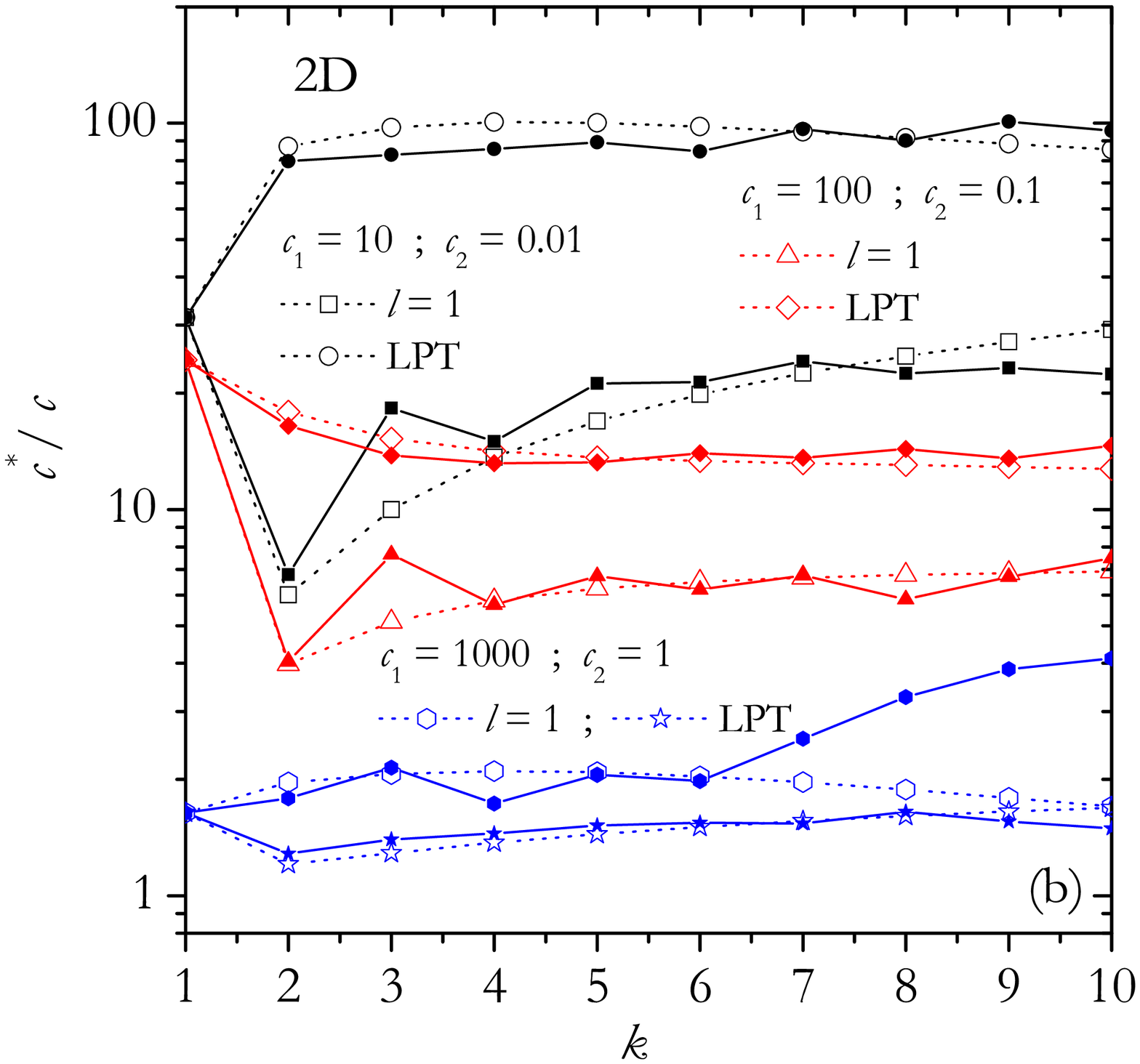}
\caption{\label{figure13} Results of the BET plots for the
adsorption in 2D heterogeneous surfaces with $l=1$ and LPT. Open
(full) symbols show results obtained from the analytic (MC)
isotherms. Dependence on $k$ of the fractions (a) $v_\mathrm{m}^*
/ v_\mathrm{m}$ and (b) $c^* / c$, for three different values of
$c_1$ and $c_2$ (being $c_1 / c_2 =10^3$) as indicated.}
\end{figure}

Finally, Figures~\ref{figure12}a and b, and
Figures~\ref{figure13}a and b show the results of the 2D BET plots
for $c_1 / c_2 =2.5$ and $c_1 / c_2 =10^3$, respectively. In all
cases, we have analyzed both analytic and MC isotherms. As we can
see, the behavior is similar to the 1D case. Nevertheless, even
taking very different values of the parameters $c_1$ and $c_2$, it
is not possible to obtain a complete compensation effect.

\section{Conclusions}

In the present paper, an analytic isotherm for the multilayer
adsorption of polyatomic molecules on different surfaces has been
proposed. The formalism reproduces the classical BET theory
\cite{Brunauer1938} and the recently reported dimer equations;
\cite{Riccardo2002} leads to the exact solution for a 1D
homogeneous substrate; and, as is demonstrated from comparison
with MC simulation, provides a good approximation for 1D
heterogeneous surfaces. With respect to 2D substrates (homogeneous
and heterogeneous surfaces), the approach is not exact. However,
MC data shows that, for molecules of moderate size (not larger
than $k=6$), the analytic isotherm behaves qualitatively similar
to the simulation.

In addition, we carry out numerical experiments to determine, in
different adsorption situations, how much the  value of the
monolayer volume predicted by the BET equation differs from its
real value. For this purpose, analytic isotherms and MC data were
analyzed as experimental data. For 1D and 2D homogeneous surfaces,
the monolayer volume calculated by the BET plots is approximately
10-30 per cent smaller than the real value. On the other hand, in
all cases, the parameter $c^*$ is always larger than $c$. As the
multilayer adsorption occurs on a bivariate heterogeneous surface,
a compensation effect is found, with very different values of $c$.
Nevertheless, in any of the considered cases, this compensation is
not enough to eliminate the decrease caused by the molecular size.

\section*{Acknowledgments}

This work was supported in part by CONICET (Argentina) under
project PIP 6294; Universidad Nacional de San Luis (Argentina)
under project 322000; Universidad Tecnol\'ogica Nacional, Facultad
Regional San Rafael (Argentina) under project PID PQCO SR 563 and
the National Agency of Scientific and Technological Promotion
(Argentina) under project 33328 PICT 2005.

\end{document}